\documentclass[aps,prb,twocolumn,amsmath,amssymb,floatfix,superscriptaddress]{revtex4-2}
\usepackage{graphicx}% include figure files
\usepackage{bm}% bold math
\usepackage{hyperref}% add hypertext capabilities
\hypersetup{
    colorlinks,
    citecolor=black,
    filecolor=black,
    linkcolor=black,
    urlcolor=black
}
\usepackage{braket,bbold,color,empheq,enumerate,enumitem}
\usepackage{tcolorbox}
\usepackage{tabularx}
\usepackage{makecell}
\usepackage{comment}
\usepackage{lipsum}
\usepackage{dsfont}
\usepackage{tikz,xcolor}

\newcommand{\bs}[1]{\boldsymbol{#1}}

\def\z{\mathbb{Z}}		
\def\ztwo{\mathbb{Z}_2}

\begin{document}

\title{Infernal and Exceptional Edge Modes: \texorpdfstring{\\}{}Non-Hermitian Topology Beyond the Skin Effect}

\author{M. Michael Denner}
\affiliation{Department of Physics, University of Zürich, Winterthurerstrasse 190, 8057, Zürich, Switzerland}

\author{Titus Neupert}
\affiliation{Department of Physics, University of Zürich, Winterthurerstrasse 190, 8057, Zürich, Switzerland}

\author{Frank Schindler}
\affiliation{Blackett Laboratory, Imperial College London, London SW7 2AZ, United Kingdom}

\begin{abstract}
The classification of point gap topology in all local non-Hermitian symmetry classes has been recently established. However, many entries in the resulting periodic table have only been discussed in a formal setting and still lack a physical interpretation in terms of their bulk-boundary correspondence. Here, we derive the edge signatures of all two-dimensional phases with intrinsic point gap topology. While in one dimension point gap topology invariably leads to the non-Hermitian skin effect, non-Hermitian boundary physics is significantly richer in two dimensions. We find two broad classes of non-Hermitian edge states: (1)~\emph{Infernal points}, where a skin effect occurs only at a single edge momentum, while all other edge momenta are devoid of edge states. Under semi-infinite boundary conditions, the point gap thereby closes completely, but only at a single edge momentum. (2)~Non-Hermitian exceptional point \emph{dispersions}, where edge states persist at all edge momenta and furnish an anomalous number of symmetry-protected exceptional points. Surprisingly, the latter class of systems allows for a finite, non-extensive number of edge states with a well defined dispersion along all generic edge terminations. Instead, the point gap only closes along the real and imaginary eigenvalue axes, realizing a novel form of non-Hermitian spectral flow.
\end{abstract}

\maketitle

\section{Introduction} \label{sec: intro}

The study of non-Hermitian (NH) band theory has gained increasing attention in recent years, with potential applications in optics~\cite{Feng:2017,Midya:2018,Miri:2019, Gong:2022:2,Roccati:2022,Roccati:2023}, condensed matter physics~\cite{Kozii:2017,Yoshida:2018,Shen:2018,Budich:2019:symmetry,Herviou:2019:entanglement,Aguado:2019,Hamazaki:2019:many, Michishita2020, Michishita2020-2, Nagai2020,Matsumoto:2020:continuous,Hofmann20,Chang:2020:entanglement,Vecsei21,Zhang:2021,Schindler21NHDefects,Liu:2021:supermetal,Kawabata:2021:nonunitary,Sayyad:2022:3,Kawabata:2022:many,Guo:2023:boost,Fulga22,Schindler23}, and quantum information processing~\cite{Ju:2019,Gong:2022,Fleckenstein:2022,Kawabata:2022,Xiao:2022:level,Kawabata:2022:entanglement}. NH topological phases have attracted particular interest due to their unconventional bulk-boundary correspondence~\cite{Kawabata:2018:anomalous,FoaTorres2019,Edvardsson:2019:extensions,Kunst:2019:systems,Kawabata:2019,Liu:2019:second,Kawabata:2019:topo,Herviou:2019:defining,Ashida2020,Edvardsson:2020:phase,Kawabata:2020:real,Hamazaki:2020:universality,Lee:2020:unraveling,Kawabata:2020:nonbloch,Bergholtz2021,Koch2020,Xiao2020,eti,Kawabata2021,Stegmaier:2021:topo,Vyas2021,Lee2019_periodic,Lee:2019-2,Kawabata:2020:higher}. One of the prime examples is the NH skin effect, in which an extensive number of states localizes at the boundary of a one-dimensional (1D) system~\cite{Xiong:2018,Wang:2018,Kunst:2018,Torres:2018,Lee2019_skineffect,Helbig2020,Sato-PRL:2020,Weidemann311,Borgnia2020,Okuma2020}. The NH skin effect can be enhanced by symmetries, for example to a $\ztwo$ skin effect in 1D time reversal-symmetric systems~\cite{Sato-PRL:2020}. However, NH topological phases are not limited to 1D: NH internal symmetries give rise to a total of 38 symmetry classes, which were topologically classified for all spatial dimensions in the seminal works of Refs.~\onlinecite{Ueda:2018,Kawabata:2019,Sato-PRL:2020}. The physical consequence of this classification are new forms of dynamical and topologically protected edge states. Intrinsically NH systems have a point-gapped bulk spectrum, in which complex eigenvalues surround a region devoid of eigenstates~\cite{Kawabata:2019,Bergholtz2021}. Crucially, such Hamiltonians cannot be adiabatically deformed to purely (anti-) Hermitian limits. While the NH skin effect constitutes the bulk-boundary correspondence of nontrivial 1D point gap topology, to date, there is no systematic study of the boundary physics of two-dimensional (2D) NH systems.

\begin{figure*}[t]
\centering
\includegraphics[width=1\textwidth,page=1]{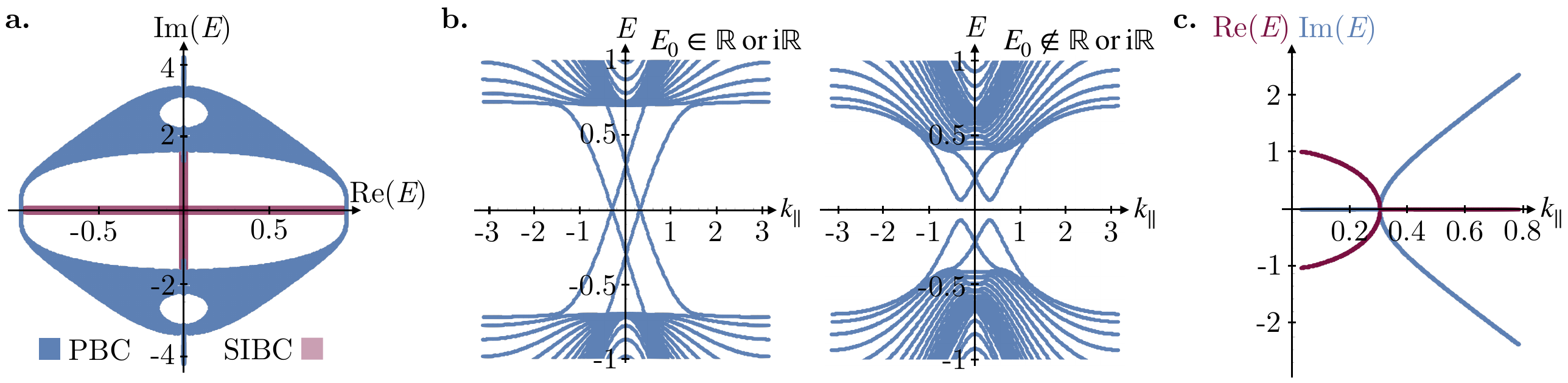}\vspace{-2mm}
\caption{\textbf{Edge exceptional points.} \textbf{a.}~Spectrum in the complex plane under PBC (blue) and SIBC (blue and red) for a model exhibiting nontrivial point gap topology in NH class AIII$^{S_{-}}$~\cite{Note1}. Boundary-localized states fill the SIBC spectral point gap along the real and imaginary axis. \textbf{b.}~The corresponding EHH OBC spectrum for $E_0$ along the real and imaginary axis inside the point gap contains one left- and one right-moving chiral mode per edge (left) and gap for generic complex  $E_0 \in \mathbb{C}$ (right). The momenta at which these modes cross zero energy move as a function of $E_0$, corresponding to a dispersing NH edge state via Eq.~\eqref{eq:HermitianDoubleDefinition}. \textbf{c.}~Real (red) and imaginary (blue) dispersion of the corresponding NH edge state in OBC, exhibiting a single exceptional point. The bulk states are not shown for clarity.\label{fig:AIIISm}}
\end{figure*}

In this paper, we derive the edge signature of all intrinsically point-gapped phases in 2D. These phases cannot be trivialized by coupling to NH line-gapped phases or symmetry-preserving perturbations as long as the point gap remains open~\cite{Sato-PRL:2020}. To establish the bulk-boundary correspondence, we focus on the spectrum with semi-infinite boundary conditions (SIBC). The SIBC spectrum is related to the $\epsilon$-pseudospectrum, which captures the behavior of a Hamiltonian under small perturbations of order $\mathcal{O}(\epsilon)$~\cite{Sato-PRL:2020,Okuma2020}. As such, it provides a robust observable for NH systems, which in general can be highly sensitive to infinitesimal errors. Moreover, only the SIBC spectrum -- and not the spectrum under open boundary conditions (OBC) -- allows for a one-to-one correspondence between the bulk topological invariants, calculated in periodic boundary conditions (PBC), and boundary dispersion~\cite{fluxes}. Our approach, therefore, differs from the OBC treatment of Ref.~\onlinecite{Nakamura:2022}. Depending on the NH symmetry class, the boundary response falls into one of two classes:
\begin{enumerate}[label=\textbf{(\arabic*)}]
    \item{\textbf{Infernal points (IPs):} The point gap remains open at generic values of the edge momentum $k_\parallel$, but completely fills up with an extensive number of SIBC edge states at one of $k_\parallel = 0,\pi$ (see Fig.~\ref{fig:AIIdag}a).}
    \item{\textbf{Exceptional points (EPs):} As $k_\parallel$ is varied, the edge state disperses with a square-root singularity for real and imaginary parts of the spectrum~\cite{Kawabata:2019:3,Sayyad:2022,Sayyad:2022:2}. Concomitantly, the SIBC point gap only fills up along the real and imaginary axes (see Fig.~\ref{fig:AIIISm}a).}
\end{enumerate}
Both types of edge states are topologically protected and anomalous in the sense that they cannot be realized in a 1D lattice system. We begin by discussing case (1) for the specific example of NH symmetry class AII$^\dagger$, highlighting the unique spectral signature and edge state. Next, we derive the general criterion for infernal edge modes. As an example for a NH symmetry class without an IP, we then discuss case (2) in NH symmetry class AIII$^{S_{-}}$. Based on these two paradigmatic signatures, we identify the nature of edge state for each NH symmetry class that allows for intrinsic point gap topology in Tab.~\ref{tab:sym_classes}.

\section{Infernal edge modes}
\begin{figure*}
\centering
\includegraphics[width=1\textwidth,page=1]{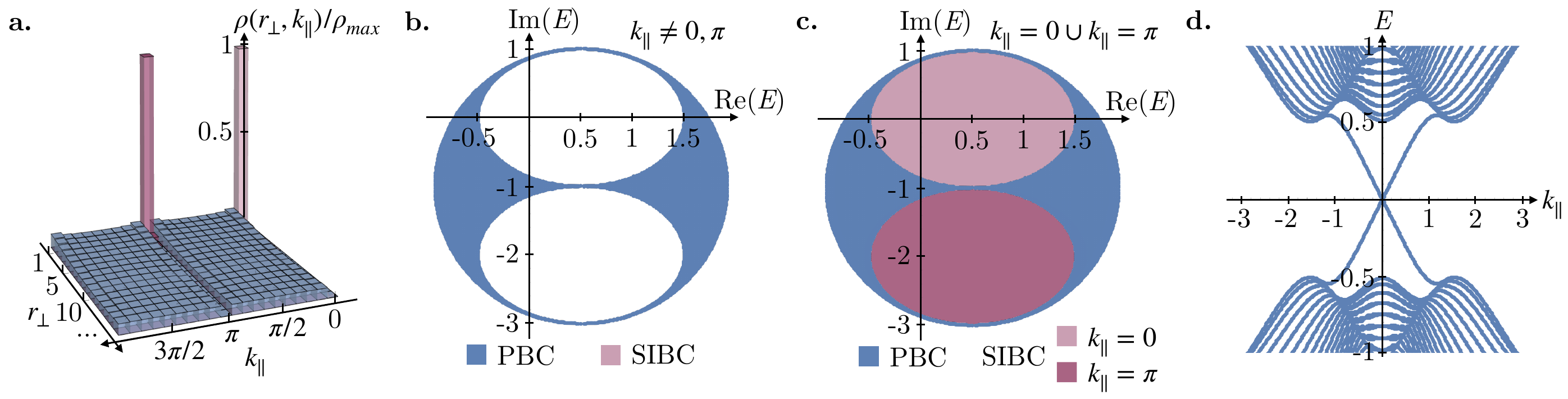}\vspace{-2mm}
\caption{\textbf{Edge infernal points.} \textbf{a.}~Infernal point mode localization in hybrid real and momentum space for a model with nontrivial point gap topology in NH symmetry class AII$^\dagger$~\cite{Note1}. An extensive number of eigenmodes accumulate at the boundary of the system for edge momentum $k_\parallel = 0,\pi$, indicated by a peak of the summed density $\rho(r_\perp,k_\parallel) = \sum_{\alpha, i} |\langle r_{\perp,i} | \psi_\alpha(k_\parallel)\rangle |^2$, where $\alpha$ ranges over \emph{all} eigenstates $\ket{\psi_\alpha(k_\parallel)}$ of the OBC Hamiltonian, $r_\perp$ is the lattice coordinate perpendicular to the edge, and $i$ runs over all sublattice and orbital degrees of freedom. This system has two nontrivial point gaps in PBC, one of which contributes the peak at $k_\parallel = 0$ while the other contributes the peak at $k_\parallel = \pi$. \textbf{b, c.}~Spectrum in the complex plane under PBC (blue) and SIBC (blue and red). Boundary-localized states fill the SIBC spectral point gap at a single boundary momentum $k_\parallel = 0$ for the upper point gap and $k_\parallel = \pi$ for the lower point gap. Bulk states are depicted for all momenta in panels b and c. \textbf{d.}~For any eigenvalue $E_0$ inside the point gap around $E_0=0$, the EHH OBC spectrum exhibits one helical zeromode per edge at $k_\parallel = 0$. \label{fig:AIIdag}}
\end{figure*}

We begin our survey of edge modes with the specific example of NH symmetry class AII$^\dagger$, which was also discussed in the appendix of Ref.~\onlinecite{Sato-PRL:2020}. This class is characterized by a pseudo-time-reversal symmetry $U_{\mathcal{T}} \mathcal{H}(\bs{k})^TU_{\mathcal{T}}^\dagger = \mathcal{H}(-\bs{k})$, where $U_{\mathcal{T}}$ is a unitary operator obeying $U_{\mathcal{T}}U_{\mathcal{T}}^* = -1$. The intrinsic point gap topology is classified by a $\ztwo$ invariant $\nu(E_0) \in \{0,1\}$~\cite{Kawabata:2019}, where $E_0$ is any eigenvalue inside the point gap. Relying on the topological equivalence between a NH Hamiltonian $H$ and an extended Hermitian Hamiltonian (EHH) $\bar{H}$, defined as~\cite{Mudry1998,Mudry1998-2}
\begin{equation} 
\label{eq:HermitianDoubleDefinition}
\bar{H} = \begin{pmatrix} 0 & H - E_0 \\ H^\dagger - E_0^* & 0 \end{pmatrix},
\end{equation}
we can derive the signature of the nontrivial point gap phase with $\nu=1$. The presence of a point gap of $H$ around $E=E_0$ results in a gapped spectrum of $\bar{H}$ around zero energy. (We only refer to the eigenvalues of $\bar{H}$ as energies because the eigenvalues of the NH Hamiltonian $H$ involve not only energies but also lifetimes.) On the other hand, the existence of exact topological zero-energy eigenvalues in $\bar{H}$ corresponds to protected states at $E=E_0$ within the NH point gap. The EHH for NH class AII$^\dagger$ is in Hermitian symmetry class DIII, \emph{regardless} of the choice of $E_0$. Class DIII Hermitian systems are also $\ztwo$-classified in 2D~\cite{Ryu2010}, so that a nontrivial point gap in NH class AII$^\dagger$ maps to a topological superconductor phase in class DIII. This phase hosts helical boundary modes that cross zero energy at a time-reversal invariant momentum (TRIM) (see Fig.~\ref{fig:AIIdag}d). Crucially, due to the time-reversal (TRS), particle-hole (PHS) and chiral symmetry (CS) of Hermitian class DIII, the Kramers pair cannot be moved away from the TRIM and zero energy. In the NH SIBC spectrum, this zeromode corresponds to a single edge-localized state at complex eigenvalue $E_0$ in the point gap (see Fig.~\ref{fig:AIIdag}c)~\cite{Sato-PRL:2020}. By repeating this construction for all $E_0$ within the point gap, we obtain a number of modes localized at the boundary that scales with the linear system size. However, since the helical crossing of the EHH edge mode cannot move away from the TRIM even as $E_0$ is varied, this extensive accumulation of modes appears only for a single edge momentum $k_\parallel \in \{0,\pi\}$ (see Fig.~\ref{fig:AIIdag}a). For \emph{all} other momenta, the point gap remains empty (see Fig.~\ref{fig:AIIdag}b). This dispersion with $k_\parallel$ constitutes an intrinsically NH edge state that we dub \emph{infernal point} (IP) in reminiscence of the infinitely steep surface dispersion observed in some 3D NH point-gapped systems~\cite{eti}. (See the Supplemental Material~\footnote{The Supplemental Material contains exhaustive auxiliary derivations and toy model Hamiltonians realising intrinsic point gap topology.} for a discussion of the OBC and $\epsilon$-pseudospectrum associated with an IP.)

\begin{table}[t]
\centering
\begin{tabular}{ l | c c c | w{c}{3.5em} w{c}{3.5em} | c }
\hline \hline
Class & \multicolumn{3}{c|}{Symmetry} & \multicolumn{2}{c|}{\makebox[2pt]{Classification}} & Edge mode\\
& TRS$^{(\dagger)}$ & PHS$^{(\dagger)}$ & CS & 2D & 1D  &\\
\hline \hline
AII$^\dagger$ & -1 & - & -& $\ztwo$ & $\ztwo$  & IP (1x)\\
DIII$^\dagger$ & -1 & +1 & 1 & $\ztwo$ & $\ztwo$ & IP (1x)\\
BDI$^{S_{+-}}$ & +1 & +1 & 1& $\ztwo$ &$\ztwo$ &IP (1x)\\
D$^{S_{-}}$ & - & +1 & - &$\ztwo$ &$\ztwo$& IP (1x)\\
AIII$^{S_{-}}$ & - & -& 1& $\ztwo$& 0& EP (1x)\\
DIII$^{S_{+-}}$ & -1 & +1 & 1& $\ztwo$ &0 &EP (2x)\\
CII$^{S_{-+}}$ & -1 & -1 & 1&$\ztwo$&  0& EP (2x)\\
CII$^{S_{+-}}$ & -1 & -1 & 1&$\ztwo$ &0 &EP (2x)\\ 
CI$^{S_{-+}}$ & +1 & -1 & 1 &$\ztwo$ &0 &EP (2x)\\
\hline \hline
\end{tabular}
\caption{\label{tab:sym_classes} \textbf{NH bulk-boundary correspondence in 2D.} We list all NH symmetry classes with intrinsic point gap topology~\cite{Kawabata:2019}. The symmetry class labels refer to a given NH Altland-Zirnbauer (AZ) class or its AZ$^\dagger$ counterpart, in some cases supplemented by an additional sublattice symmetry (SLS) $S$, which for NH systems is different from the AZ chiral symmetry (CS)~\cite{Kawabata:2019}. The subscript of SLS determines whether SLS commutes (+) or anticommutes (-) with the respective AZ symmetry. If both time-reversal symmetry (TRS) and particle-hole symmetry (PHS) are present, the first subscript refers to TRS and the second one to PHS. The symmetry classes are identified by the square of TRS, acting as $U_{\mathcal{T}} \mathcal{H}(\bs{k})^*U_{\mathcal{T}}^\dagger = \mathcal{H}(-\bs{k})$, and PHS, defined by $U_{\mathcal{P}}\mathcal{H}(\bs{k})^T U_{\mathcal{P}}^\dagger = - \mathcal{H}(-\bs{k})$, or TRS$^{\dagger}$, denoted by $U_{\mathcal{T}} \mathcal{H}(\bs{k})^TU_{\mathcal{T}}^\dagger = \mathcal{H}(-\bs{k})$, and PHS$^{\dagger}$, expressed as $U_{\mathcal{P}}\mathcal{H}(\bs{k})^* U_{\mathcal{P}}^\dagger = - \mathcal{H}(-\bs{k})$ (second column). Their intrinsic point gap classification (third column) is reproduced from Ref.~\onlinecite{Sato-PRL:2020}. The edge states are identified as either an infernal point (IP) or an anomalous number of exceptional points (EP) per edge and point gap. Their multiplicity is denoted in parentheses.}
\end{table}

The derivation outlined above is completely general and does not rely on a particular model Hamiltonian, but only the respective symmetry class as well as the presence of a topologically nontrivial point gap. In order to further rationalize the IP, we now consider a representative Hamiltonian for class AII$^\dagger$ and solve for the edge state within a Dirac approximation. The full Hamiltonian is given in the Supplemental Material~\cite{Note1}, for our purposes it is enough to consider its expansion to first order around $\bs{k} = 0$: 
\begin{equation}
\mathcal{H}(\bs{k}) = \left[\mathrm{i} m + \Delta\right] \sigma_0 + k_x \sigma_x - k_y \sigma_z.
\label{eq:DiracH-AIIdag}
\end{equation}
Here, $m$ and $\Delta$ are real parameters, $\sigma_{\mu}$ are the Pauli matrices ($\mu = 0, x, y, z$), and $U_{\mathcal{T}} = \mathrm{i} \sigma_y$. 

We want to stress that the full bulk Hamiltonian of this model~\cite{Note1} includes non-trivial non-Hermitian momentum dependent terms not present in its Dirac expansion. Therefore, this Hamiltonian does not simply result from an imaginary shift of a Hermitian model. In fact, a simple imaginary shift of a Hermitian topological insulator always results in a line-gapped non-Hermitian model. Here we focus on intrinsically point-gapped non-Hermitian systems to exclude such a scenario. 

To obtain the edge states of $\mathcal{H}(\bs{k})$, we consider open boundary conditions in $x$-direction, and model the transition to the surrounding vacuum by a domain wall in $m$, whose sign governs the bulk NH topology~\cite{Sato-PRL:2020,Schindler:2020}. Consequently we solve
\begin{equation}
\left\{ [\mathrm{i} m(x) + \Delta] \sigma_0 -i \partial_x \sigma_x - k_y \sigma_z \right\} \psi(x) = E \psi(x),
\end{equation}
with $E \in \mathbb{C}$, for the right eigenfunction $\psi(x)$. Here, the mass $m(x) = m_0 \mathrm{sgn}(x)$, $m_0>0$, changes sign at $x = 0$. Defining 
\begin{equation}
\omega_+^2 = \underbrace{[m_0+\mathrm{i} (k_y +E-\Delta)]}_{\alpha_+}\underbrace{[m_0-\mathrm{i} (k_y -E+\Delta)]}_{\alpha_-},
\end{equation}
the normalizable solution in the right half plane (+) reads
\begin{equation}
\psi_+(x) = \frac{1}{\mathcal{N}} e^{-\omega_+ x} \left[\frac{\omega_+}{\alpha_+}, -1 \right]^{\mathrm{T}},
\end{equation}
where $\mathcal{N}$ is a normalization constant.
Similarly, defining
\begin{equation}
\omega_-^2 = \underbrace{[m_0+\mathrm{i} (k_y -E+\Delta)]}_{\beta_+}\underbrace{[m_0-\mathrm{i} (k_y +E-\Delta)]}_{ \beta_-},
\end{equation}
the normalizable solution in the left halfspace (-) reads
\begin{equation}
\psi_-(x) = \frac{1}{\mathcal{N}} e^{\omega_- x} \left[\frac{\omega_-}{\beta_-}, -1 \right]^{\mathrm{T}}.
\end{equation} 
Matching the solutions across the domain wall yields
\begin{equation}
\psi_{+} (0) = \psi_{-} (0) \quad \Leftrightarrow \quad \frac{\alpha_-}{\alpha_+}=\frac{\beta_+}{\beta_-}.
\label{eq:matching-condition}
\end{equation}
We now have to distinguish three cases: first, when neither $\beta_-$ nor $\alpha_+$ are equal to zero, the matching constraint implies $k_y = 0$. However, all possible (real and imaginary) energies $E$ are allowed. Furthermore, the case $\beta_- = 0$ is not admissible because the solution should decay in the vacuum. In the bulk however, this case might be tolerated, so we can consider $\alpha_+ = 0$. From Eq.~\eqref{eq:matching-condition}, finite $\beta_{\pm} \neq 0$ additionally implies $\alpha_- = 0$, again resulting in $k_y = 0$. Consequently, away from $k_y = 0$ no edge state solution exists. At $k_y = 0$, however, we obtain infinitely many boundary modes, with arbitrary $E \in \mathbb{C}$. Concomitantly, the entire NH point gap fills with edge-localized states at a single momentum, the defining characteristic of an IP. This result holds irrespective of the specific model Hamiltonian used above as long as the point gap is not closed and the symmetries of class AII$^\dagger$ are preserved: see the Supplemental Material~\cite{Note1} for an explicit study on the persistence of the IP under generic symmetry-allowed perturbations to Eq.~\eqref{eq:DiracH-AIIdag} respecting the point gap topology, \emph{i.e.}, they do not close or close and reopen the point gap, including traceless Pauli matrices with imaginary and $k$-dependent prefactors.
We can also calculate the left eigenstates of $\mathcal{H}(\bs{k})$ in Eq.~\eqref{eq:DiracH-AIIdag}, which are the right eigenstates of $\mathcal{H}(\bs{k})^\dagger$. The overlap between left and right eigenstates vanishes, similar to the NH skin effect in 1D~\cite{Sato-PRL:2020}.

Notably, an IP cannot be realized in a 1D lattice system with a finite-dimensional unit cell Hilbert space. Instead, such an edge mode realizes an anomalous dispersion whose discontinuity and TRIM skin effect capitalizes on a topologically nontrivial 2D bulk. To generalize beyond the specific example given above, we investigate the presence of infernal edge modes in all NH symmetry classes with intrinsic point gap topology. We find that a point gap-nontrivial 2D system in a given NH symmetry class exhibits an IP \emph{if and only if} this symmetry class also has a nontrivial 1D point gap classification~\cite{Note1}. The NH symmetry classes satisfying this criterion are summarized in Tab.~\ref{tab:sym_classes}.

\section{Exceptional edge modes}
We next consider the example of NH symmetry class AIII$^{S_{-}}$, which contains a CS $U_{\mathcal{C}}\mathcal{H}(\bs{k}) U_{\mathcal{C}}^\dagger = - \mathcal{H}(\bs{k})^\dagger$ as well as a sublattice symmetry $\mathcal{S}\mathcal{H}(\bs{k}) \mathcal{S}^\dagger = - \mathcal{H}(\bs{k})$ with $\{U_{\mathcal{C}} , \mathcal{S}\} = 0$~\cite{Kawabata:2019}. This symmetry class quantizes the 2D point gap topological invariant $C_1\in \mathbb{Z}$~\cite{Kawabata:2019}. However, the classification of intrinsic point gap topology in AIII$^{S_{-}}$ is only $\mathbb{Z}_2$, corresponding to $C_1 \mod 2$: all phases with even $C_1$ can be trivialized by coupling to line-gapped phases~\cite{Sato-PRL:2020}. To derive the topological edge state of the phase where $|C_1|=1$, we again rely on the EHH for a given eigenvalue $E_0$ inside the point gap [Eq.~\eqref{eq:HermitianDoubleDefinition}]. Importantly, and unlike for symmetry classes protecting infernal edge modes, the EHH enjoys distinct Hermitian symmetries depending on the choice of $E_0$~\cite{Note1}. For purely real ($E_0 \in \mathbb{R}$) or imaginary energies ($E_0 \in \mathrm{i}\mathbb{R}$), the EHH can be block-diagonalized into two matrix blocks that are mapped to each other under CS (see below for details)~\cite{Note1}. Each block individually only satisfies Hermitian class A, which is $\z$-classified in 2D~\cite{Ryu2010}. The EHH associated with $|C_1| = 1$ then corresponds to an insulator with Chern number $C=\pm 1$ in each of the two blocks, giving rise to a total of one right and one left-moving chiral mode per edge. These modes are protected from gapping out due to the symmetries present for $E_0 \in \mathbb{R}$ ($E_0 \in \mathrm{i}\mathbb{R}$), and cross zero energy at distinct edge momenta depending on the particular choice of $E_0$ (see Fig.~\ref{fig:AIIISm}b, left). The corresponding NH SIBC spectrum therefore exhibits edge states dispersing as a function of the edge momentum. However, the edge states cross the point gap only along the real and imaginary axis (see Fig.~\ref{fig:AIIISm}a). Away from these axes, the SIBC spectrum shows no in-gap states, because the EHH symmetries reduce for generic $E_0 \in \mathbb{C}$ and do not anymore prevent hybridization between the two chiral edge modes~\cite{Note1,Ryu2010} (see Fig.~\ref{fig:AIIISm}b, right). We note that even though the NH edge states only traverse the point gap along special axes, they cannot be moved away from these axes or out of the point gap due to the symmetry of NH class AIII$^{S_{-}}$.

\begin{figure*}[t]
\centering
\includegraphics[width=1\textwidth,page=1]{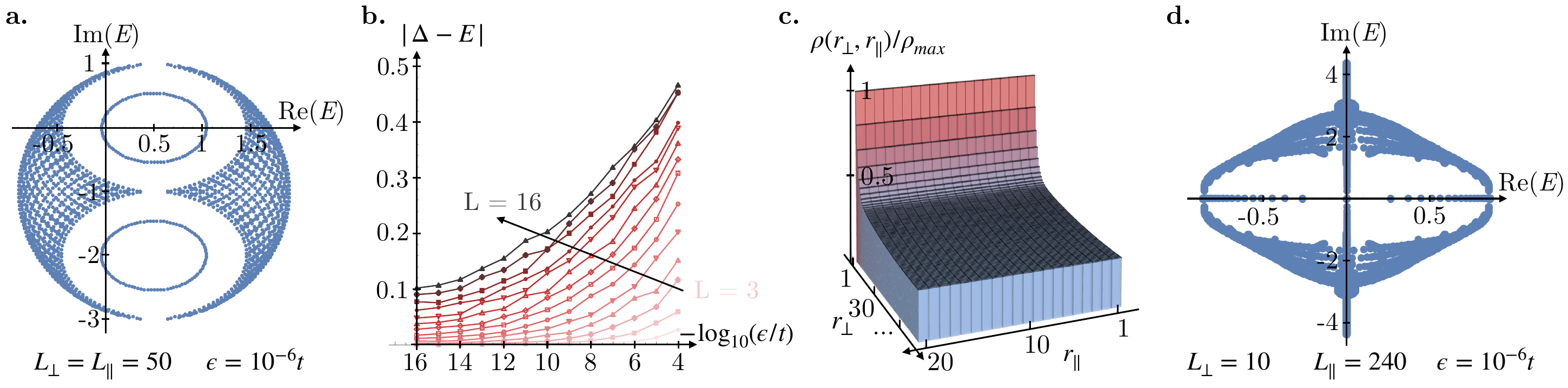}\vspace{-2mm}
\caption{\textbf{Experimental signatures.} \textbf{a.}~Spectrum under OBC in $r_\perp$-direction ($L_\perp = 50$) and PBC in $r_\parallel$-direction ($L_\parallel = 50$) for the infernal point model of Fig.~\ref{fig:AIIdag} in presence of disorder with $\epsilon = 10^{-6} t$ (Sec.~\ref{sec: exprel}), where $t$ is the mean hopping strength~\cite{Note1}. In contrast to the SIBC spectrum where the point gap is fully covered by edge-localized states, the infernal point manifests as a ellipse of eigenvalues in the complex plane. Its minimal distance from the point gap center -- the separation of the ``perihelion" from $E = \Delta$~\cite{Note1} -- increases with $L_\perp L_\parallel$ and $\epsilon$, as shown in panel \textbf{b} (where $L \equiv L_\perp = L_\parallel$). \textbf{c.}~Infernal point localization in real space. Notwithstanding the presence of disorder, an extensive number of eigenmodes accumulate at the boundary of the system in $r_\perp$-direction, indicated by a peak of the summed density $\rho(r_\perp, r_\parallel) = \sum_{\alpha, i} |\langle r_\perp, r_\parallel, i | \psi_\alpha\rangle |^2$, where $\alpha$ ranges over \emph{all} eigenstates $\ket{\psi_\alpha}$ of the real-space Hamiltonian, and $i$ runs over all degrees of freedom within the unit cell. \textbf{d.}~Spectrum under OBC in $r_\perp$-direction ($L_\perp = 10$) and PBC in $r_\parallel$-direction ($L_\parallel = 240$) for the exceptional point model of Fig.~\ref{fig:AIIISm} in presence of disorder with $\epsilon = 10^{-6} t$ (Sec.~\ref{sec: exprel}), where $t$ is the mean hopping strength~\cite{Note1}. Unlike for the infernal point, this spectrum approaches the corresponding SIBC spectrum (Fig.~\ref{fig:AIIISm}) in the thermodynamic limit even at finite disorder. This observation remains valid for both larger system sizes and $\epsilon > 10^{-6}$.\label{fig:experiment}}
\end{figure*}

We now show that the NH edge dispersion realizes a single EP (also evident from the numerical result of Fig.~\ref{fig:AIIISm}c). Due to the absence of a NH skin effect at all edge momenta, we can derive the edge state by writing down the effective edge EHH and then inverting Eq.~\eqref{eq:HermitianDoubleDefinition} to obtain the effective edge NH Hamiltonian.
At $E_0 = 0$, the EHH for NH class AIII$^{S_{-}}$ features three \emph{distinct} chiral symmetries, $\bar{U}_{\mathcal{C}}$, $\bar{\Sigma}_{\mathcal{C}}$, and $\bar{\mathcal{S}}$ that satisfy the algebra $\{\bar{U}_{\mathcal{C}},\bar{\Sigma}_{\mathcal{C}}\} = 0$, $\{\bar{U}_{\mathcal{C}},\bar{\mathcal{S}}\} = 0$, and $[\bar{\Sigma}_{\mathcal{C}},\bar{\mathcal{S}}] = 0$~\cite{Note1}. The minimal EHH matrix dimension that realizes this algebra is $4$. Using the Pauli matrices  $\tau_{\mu},\sigma_{\mu}$ ($\mu = 0, x, y, z$), one possible representation is given by $\bar{U}_{\mathcal{C}} = \tau_x \sigma_z, \bar{\Sigma}_{\mathcal{C}} = \tau_z \sigma_0$, and $\bar{\mathcal{S}} = \tau_0 \sigma_x$, which conforms with the basis choice in Eq.~\eqref{eq:HermitianDoubleDefinition} where $\bar{\Sigma}_{\mathcal{C}}$ is diagonal. We now combine the three chiral symmetries to form two commuting unitary symmetries $\bar{U}_1 = \bar{U}_{\mathcal{C}}\bar{\mathcal{S}}$ and $\bar{U}_2 = \bar{U}_{\mathcal{C}}\bar{\Sigma}_{\mathcal{C}}$ that both square to $-1$. Correspondingly, at $E_0 = 0$, all eigenstates of the effective edge EHH $\bar{\mathcal{H}}_{\mathrm{edge}}(k)$ are labelled by a pair of quantum numbers $(\pm \mathrm{i},\pm \mathrm{i})$ that denote the eigenvalues of $\bar{U}_1$ and $\bar{U}_2$, respectively. Any of the three chiral symmetries anti-commute with both unitary symmetries, and therefore flip \emph{both} their eigenvalues. In the common eigenbasis of $\bar{U}_1$ and $\bar{U}_2$ given by $\{\ket{\mathrm{i},\mathrm{i}},\ket{-\mathrm{i},-\mathrm{i}},\ket{\mathrm{i},-\mathrm{i}},\ket{-\mathrm{i},\mathrm{i}}\}$, we therefore posit
\begin{equation}
    \bar{\mathcal{H}}_{\mathrm{edge}}(k) = \begin{pmatrix}
        v k \sigma_z & 0 \\
        0 & \Delta \sigma_z \\
    \end{pmatrix},
    \label{eq:HermEdgeH}
\end{equation}
where $v$ is the Fermi velocity of the chiral modes and $\Delta > 0$ is a constant energy. Without loss of generality, this Hamiltonian realizes a single pair of left- and right-moving chiral modes together with a pair of gapped bands at energies $\pm \Delta$ that are necessary to conform with the symmetry realization chosen above. To satisfy the chiral symmetries, the two chiral modes appear with opposite eigenvalues of $\bar{U}_1$ and $\bar{U}_2$, preventing them from hybridization at all real $E_0 \in \mathbb{R}$ (where only $\bar{U}_1$ survives), and all imaginary $E_0 \in \mathrm{i} \mathbb{R}$ (where only $\bar{U}_2$ survives), but not at generic $E_0 \in \mathbb{C}$.
To obtain the corresponding NH Hamiltonian, we transform Eq.~\eqref{eq:HermEdgeH} back to the basis of Eq.~\eqref{eq:HermitianDoubleDefinition} in which the common eigenvectors of $\bar{U}_1$ and $\bar{U}_2$ read $\psi_{(\lambda_1 \mathrm{i},\lambda_2 \mathrm{i})} = (1, \lambda_1 \lambda_2, -\lambda_2 \mathrm{i}, \lambda_1 \mathrm{i})^{\mathrm{T}}/2$ for $\lambda_{1,2} = \pm 1$. The transformed matrix then assumes the canonical EHH form of Eq.~\eqref{eq:HermitianDoubleDefinition}, from which one can extract the effective NH edge Hamiltonian
\begin{equation}
    \mathcal{H}_{\mathrm{edge}}(k) = \frac{\mathrm{i}}{2} (v k - \Delta) \, \sigma_z + \frac{1}{2} (v k + \Delta) \, \sigma_y.
    \label{eq:EP-Hamiltonian_AIIISm}
\end{equation}
The NH spectrum $E_{\mathrm{edge}}(k) = \pm \sqrt{v k \Delta}$ indeed realizes a single EP as $k$ is varied, analogous to the numerical result of Fig.~\ref{fig:AIIISm}c. 

We note that EPs generically occur in 2D rather than 1D, because two momenta must be tuned to ensure a two-fold degeneracy between NH bands absent symmetry~\cite{Schen:2018}. However, imposing NH symmetry class AIII$^{S_{-}}$ reduces the number of degeneracy constraints to one, thereby ensuring that also in the 1D case EPs can only be annihilated in pairs~\cite{Kawabata:2019:3,Sayyad:2022,Sayyad:2022:2}. As a consequence, the presence of a single EP in Fig.~\ref{fig:AIIISm}c and Eq.~\eqref{eq:EP-Hamiltonian_AIIISm} implies an anomaly: any odd number of EPs cannot be realized in a regularized lattice system due to the NH fermion doubling theorem~\cite{Schnyder:2019:1}. 

We furthermore note that trivial and non-trivial unpaired EPs can also exist beyond the restrictions of the fermion doubling theorem as recently shown in Ref.~\onlinecite{Konig:2022}. The EPs discussed here are, however, stable to any symmetry-allowed perturbation that does not close the point gap as it is ramped up, and do not rely on a braiding of eigenvalues in 2D momentum space~\cite{Konig:2022}. Instead, they appear on the 1D boundary of a 2D system and are a direct consequence of the anomalous edge states of the corresponding EHH. Hence, edge EPs cannot be realized in any purely 1D model and correspond to an anomaly also in the non-Hermitian setting.

We investigate all 2D NH symmetry classes with intrinsic point gap topology and find that exceptional edge modes arise whenever there is no IP, as summarized in Tab.~\ref{tab:sym_classes} (see the Supplemental Material~\cite{Note1} for details). Notably, certain NH symmetry classes show two EPs on their boundary. This edge response is, however, still anomalous because 1D lattice systems in these symmetry classes can only support a multiple of four EPs.

\section{Experimental relevance} \label{sec: exprel}
In our discussion of infernal and exceptional edge modes, we have so far focussed on the SIBC spectrum because it is equal to the $\epsilon$-pseudospectrum of the corresponding OBC system~\cite{Sato-PRL:2020,Okuma2020} for an $\epsilon$ that is exponentially small in the linear system size $L$ (see also Sec.~\ref{sec: intro}). For clarity, we here only assume OBC in one direction and maintain PBC in the other, and consider full OBC later. We now study to what extent infernal and EPs are measurable at the boundaries of realistic experimental samples where $L$ is thermodynamically large but $\epsilon$ remains finite. Notably, we focus on a single-particle picture. The extensive accumulation of states at an IP violates Pauli's exclusion principle for fermionic many-body systems, and is hence modified in this context~\cite{Faisal:2022}. Bosonic or classical platforms will, however, still realize an essentially single-particle skin effect. At the same time, EPs survive even in a generic (bosonic and fermionic) many-body setting, similar to the boundary states of a Hermitian topological insulator.

We model the error $\epsilon$ by introducing uniformly distributed on-site disorder $V_\epsilon=\sum_{\bs{r},i} \delta_{\bs{r},i} \ket{\bs{r},i}\bra{\bs{r},i}$ to the respective real-space Hamiltonian~\cite{Komis22}, with $\delta_i$ drawn uniformly from the range $[-\epsilon,+\epsilon]$. Here, $\bs{r}$ ranges over all unit cells and $i$ labels intra-unit cell degrees of freedom.

We find that infernal and exceptional edge modes differ drastically in their response to small finite disorder. For IP systems with OBC along one direction, the point gap remains empty as long as the linear system size $L$ is sufficiently large and $\epsilon > 0$ is a finite $L$-independent constant (we assume a quadratic system of area $L^2$ for simplicity). This can be understood by noting that at any given $L$ and $\epsilon$, the OBC edge states form an ellipse in the complex plane (Fig.~\ref{fig:experiment}a) whose minimal radius increases with $L$ and $\epsilon$ (Fig.~\ref{fig:experiment}b). We conclude that the spectral signature of edge IPs is highly sensitive to disorder in the thermodynamic limit and difficult to observe. However, the localization profile of eigenstates prevails even when they do not cover the point gap: an extensive number of eigenstates still accumulates at the OBC edges irrespective of the presence of disorder (Fig.~\ref{fig:experiment}c). In stark contrast, we find that edge EPs are unaffected by the presence of finite disorder: their OBC spectrum hosts edge states along real and imaginary axes in the complex plane (Fig.~\ref{fig:experiment}d) and approaches the SIBC spectrum in the thermodynamic limit. Similarly, their real space localization in OBC for one and two directions is not altered (see the Supplemental Material~\cite{Note1} for the analog of Fig.~\ref{fig:experiment}c for an edge EP).

For OBC in two directions, the IP edge loses also its skin effect in addition to its in-gap states~\cite{Okuma2020}. On the other hand, the EP system remains largely unaffected in full OBC and still displays in-gap states along the real and imaginary axes~\cite{Note1}.

\section{Discussion}
We have derived the edge modes of 2D NH phases with intrinsic point gap topology, providing a physical interpretation of the classification in Refs.~\onlinecite{Ueda:2018,Kawabata:2019,Sato-PRL:2020}. We find anomalous IP and EP edge modes that drastically differ in their real-space localization profile, dispersion with parallel edge momentum, and spectral stability to finite disorder. They both capitalize on a nontrivial NH bulk, in the sense that neither of them can be realized in a 1D NH lattice model. Several questions deserve further study: Can one set up a field theory for the IP anomaly, which seems to involve an effective edge Hamiltonian $\mathcal{H}_{\mathrm{edge}}(k) \propto \delta(k)$? What is the fundamental physical difference between the types of point gap topology that induce skin effects, necessitating a SIBC treatment, and those that give rise to continuous NH dispersions, more similar in spirit to the Hermitian case and well-behaved in OBC? What are their dynamical consequences, e.g., in terms of wavepacket propagation? Are there NH edge modes that are qualitatively distinct from IPs and EPs in 3D systems or in presence of crystalline symmetries? Answering these questions is crucial for leveraging NH band topology in a realistic practical setting.
\newline
\begin{acknowledgments}
We thank Shinsei Ryu, Tom\'{a}\v{s} Bzdu\v{s}ek, Abhinav Prem and Kohei Kawabata for helpful discussions. We also thank the Kavli Institute for Theoretical Physics for hosting during completion of this work. M.M.D. acknowledges support from a Forschungskredit of the University of Zurich (Grant No. FK-22-085) and thanks the Princeton Center for Theoretical Science for hosting during some stages of this work. This project has received funding from the European Research Council (ERC) under the European Union’s Horizon 2020 research and innovation programm (ERC-StG-Neupert-757867-PARATOP). F.S. was supported by a fellowship at the Princeton Center for Theoretical Science. This research was supported in part by the National Science Foundation under Grant No. NSF PHY-1748958.
\end{acknowledgments}

\bibliography{bibliography}
\end{document}

\title{Supplementary Material for\\ ``Infernal and Exceptional Edge Modes:\\ Non-Hermitian Topology Beyond the Skin Effect"}

\author{M. Michael Denner}
\affiliation{Department of Physics, University of Zürich, Winterthurerstrasse 190, 8057, Zürich, Switzerland}

\author{Titus Neupert}
\affiliation{Department of Physics, University of Zürich, Winterthurerstrasse 190, 8057, Zürich, Switzerland}

\author{Frank Schindler}
\affiliation{Blackett Laboratory, Imperial College London, London SW7 2AZ, United Kingdom}

\maketitle

\tableofcontents

\section{Symmetries in NH systems}
This section defines symmetries in NH systems and their relation to the corresponding symmetries of the EHH. Hermitian quantities are denoted with an overline. 

\tocless\subsection{NH symmetries}{sec:nh-sym}
The Hermitian time-reversal symmetry given as
\begin{equation}
\bar{U}_{\mathcal{T}} \bar{\mathcal{H}}(\bs{k})^*\bar{U}_{\mathcal{T}}^\dagger = \bar{\mathcal{H}}(-\bs{k}),\quad \quad \bar{U}_{\mathcal{T}}\bar{U}_{\mathcal{T}}^* = \pm 1,
\end{equation}
is generalized to a NH time-reversal symmetry TRS
\begin{equation}
U_{\mathcal{T}} \mathcal{H}(\bs{k})^*U_{\mathcal{T}}^\dagger = \mathcal{H}(-\bs{k}),\quad \quad U_{\mathcal{T}}U_{\mathcal{T}}^* = \pm 1,
\end{equation}
as well as a pseudo time-reversal symmetry TRS$^{\dagger}$
\begin{equation}
U_{\mathcal{T}} \mathcal{H}(\bs{k})^TU_{\mathcal{T}}^\dagger = \mathcal{H}(-\bs{k}),\quad \quad U_{\mathcal{T}}U_{\mathcal{T}}^* = \pm 1.
\end{equation}

The Hermitian particle-hole symmetry given as
\begin{equation}
\bar{U}_{\mathcal{P}}\bar{\mathcal{H}}(\bs{k})^* \bar{U}_{\mathcal{P}}^\dagger = - \bar{\mathcal{H}}(-\bs{k}), \quad \quad \bar{U}_{\mathcal{P}}\bar{U}_{\mathcal{P}}^* = \pm 1,
\end{equation}
is generalized to a NH particle-hole symmetry PHS
\begin{equation}
U_{\mathcal{P}}\mathcal{H}(\bs{k})^T U_{\mathcal{P}}^\dagger = - \mathcal{H}(-\bs{k}), \quad \quad U_{\mathcal{P}}U_{\mathcal{P}}^* = \pm 1,
\end{equation}
as well as a pseudo particle-hole symmetry PHS$^{\dagger}$
\begin{equation}
U_{\mathcal{P}}\mathcal{H}(\bs{k})^* U_{\mathcal{P}}^\dagger = - \mathcal{H}(-\bs{k}), \quad \quad U_{\mathcal{P}}U_{\mathcal{P}}^* = \pm 1.
\end{equation}

The Hermitian chiral symmetry given as
\begin{equation}
\bar{U}_{\mathcal{C}}\bar{\mathcal{H}}(\bs{k}) \bar{U}_{\mathcal{C}}^\dagger = - \bar{\mathcal{H}}(\bs{k}),\quad \quad \bar{U}_{\mathcal{C}}^2 = 1,
\end{equation}
is generalized to a NH chiral symmetry CS
\begin{equation}
\label{eq:nonHerm-chiral-sym}
U_{\mathcal{C}}\mathcal{H}(\bs{k})^\dagger U_{\mathcal{C}}^\dagger = - \mathcal{H}(\bs{k}),\quad \quad U_{\mathcal{C}}^2 = 1.
\end{equation}

Additonally, one can have sublattice symmetry SLS, defined by
\begin{equation}
\mathcal{S}\mathcal{H}(\bs{k}) \mathcal{S}^\dagger = - \mathcal{H}(\bs{k}),\quad \quad \mathcal{S}^2 = 1,
\end{equation}
as well as pseudo-Hermiticity
\begin{equation}
\eta\mathcal{H}(\bs{k})^\dagger \eta^\dagger = \mathcal{H}(\bs{k}),\quad \quad \eta^2 = 1.
\end{equation}

\tocless\subsection{Appearance of NH symmetries in the EHH}{sec:sym-herm_ext}

The presence of symmetries for the NH Hamiltonian $\mathcal{H}(\bs{k})$ imposes constraints on the EHH $\bar{\mathcal{H}}(\bs{k})$,
\begin{equation} 
\bar{\mathcal{H}}(\bs{k}) = \begin{pmatrix} 0 & \mathcal{H}(\bs{k})-E_0 \\ \mathcal{H}^\dagger(\bs{k})-E_0^* & 0 \end{pmatrix}.
\label{eq:EHH}
\end{equation}
Specifically, for $E_0 = 0$, it holds:
\begin{equation}
 \bar{U}_{\mathcal{T}} \bar{\mathcal{H}}(\bs{k})^* \bar{U}_{\mathcal{T}}^\dagger = \bar{\mathcal{H}}(-\bs{k}),
\end{equation}
with 
\begin{equation}
\bar{U}_{\mathcal{T}} =
\begin{pmatrix}
U_{\mathcal{T}} &0 \\
0 & U_{\mathcal{T}}\\
\end{pmatrix}
\end{equation}
for TRS and 
\begin{equation}
\bar{U}_{\mathcal{T}} =
\begin{pmatrix}
0 &U_{\mathcal{T}} \\
U_{\mathcal{T}}&0\\
\end{pmatrix}
\end{equation}
for TRS$^\dagger$ of the NH Hamiltonian.
\begin{equation}
 \bar{U}_{\mathcal{P}} \bar{\mathcal{H}}(\bs{k})^* \bar{U}_{\mathcal{P}}^\dagger = -\bar{\mathcal{H}}(-\bs{k}),
\end{equation}
with 
\begin{equation}
\bar{U}_{\mathcal{P}} =
\begin{pmatrix}
0 &U_{\mathcal{P}} \\
 U_{\mathcal{P}} & 0\\
\end{pmatrix},
\end{equation}
for PHS and 
\begin{equation}
\bar{U}_{\mathcal{P}} =
\begin{pmatrix}
U_{\mathcal{P}} &0 \\
0& U_{\mathcal{P}} \\
\end{pmatrix},
\end{equation}
for PHS$^\dagger$ of the NH Hamiltonian. Additionally we have

\begin{equation}
 \bar{U}_{\mathcal{C}} \bar{\mathcal{H}}(\bs{k}) \bar{U}_{\mathcal{C}}^\dagger = -\bar{\mathcal{H}}(\bs{k}),\quad \quad  \bar{U}_{\mathcal{C}} =
\begin{pmatrix}
0 &U_{\mathcal{C}} \\
 U_{\mathcal{C}} & 0\\
\end{pmatrix},
\end{equation}

\begin{equation}
\label{eq:hermext-ssym}
 \bar{\mathcal{S}} \bar{\mathcal{H}}(\bs{k}) \bar{\mathcal{S}} ^\dagger = -\bar{\mathcal{H}}(\bs{k}),\quad \quad  \bar{S} =
\begin{pmatrix}
\mathcal{S} & 0\\
 0&\mathcal{S} \\
\end{pmatrix},
\end{equation}

\begin{equation}
 \bar{\eta} \bar{\mathcal{H}}(\bs{k})^\dagger \bar{\eta}^\dagger = \bar{\mathcal{H}}(\bs{k}),\quad \quad  \bar{\eta} =
\begin{pmatrix}
0 &\eta\\
 \eta & 0\\
\end{pmatrix}.
\end{equation}

By construction, $\bar{\mathcal{H}}(\bs{k})$ enjoys an additional chiral (sublattice) symmetry for arbitrary $E_0$:
\begin{equation} 
\label{eq:hermext-chiralsym}
\bar{\Sigma}_{\mathcal{C}}\bar{\mathcal{H}}(\bs{k}) \bar{\Sigma}_{\mathcal{C}}^\dagger = - \bar{\mathcal{H}}(\bs{k}), \quad \bar{\Sigma}_{\mathcal{C}} = \begin{pmatrix} \mathbb{1} & 0 \\ 0 & - \mathbb{1} \end{pmatrix}.
\end{equation}

Besides the chiral symmetry introduced in \eqref{eq:hermext-chiralsym}, sublattice symmetry $\mathcal{S}$ and PHS $U_{\mathcal{P}}$ allow a TRS
\begin{equation}
 \bar{U}_{\mathcal{T}} \bar{\mathcal{H}}(\bs{k})^* \bar{U}_{\mathcal{T}}^\dagger = \bar{\mathcal{H}}(-\bs{k}),
\end{equation}
with 
\begin{equation}
\bar{U}_{\mathcal{T}} =
\begin{pmatrix}
0 &\mathcal{S}U_{\mathcal{P}} \\
\mathcal{S}U_{\mathcal{P}} & 0\\
\end{pmatrix}
\label{eq:altTRS}
\end{equation}
and a PHS
\begin{equation}
 \bar{U}_{\mathcal{P}} \bar{\mathcal{H}}(\bs{k})^* \bar{U}_{\mathcal{P}}^\dagger = -\bar{\mathcal{H}}(-\bs{k}),
\end{equation}
with 
\begin{equation}
\bar{U}_{\mathcal{P}} =
\begin{pmatrix}
0 &\mathcal{S} U_{\mathcal{P}} \\
-\mathcal{S}U_{\mathcal{P}} & 0\\
\end{pmatrix}.
\label{eq:altPHS}
\end{equation}
For daggered NH symmetry classes, TRS$^\dagger$
\begin{equation}
 \bar{U}_{\mathcal{T}} \bar{\mathcal{H}}(\bs{k})^* \bar{U}_{\mathcal{T}}^\dagger = \bar{\mathcal{H}}(-\bs{k}),
\end{equation}
with
\begin{equation}
\bar{U}_{\mathcal{T}} =
\begin{pmatrix}
0 &U_{\mathcal{T}} \\
U_{\mathcal{T}}&0\\
\end{pmatrix}
\end{equation}
can be combined with \eqref{eq:hermext-chiralsym} to a PHS
\begin{equation}
 \bar{U}_{\mathcal{P}} \bar{\mathcal{H}}(\bs{k})^* \bar{U}_{\mathcal{P}}^\dagger = -\bar{\mathcal{H}}(-\bs{k}),
\end{equation}
with 
\begin{equation}
\bar{U}_{\mathcal{P}} =
\begin{pmatrix}
0 &U_{\mathcal{T}} \\
-U_{\mathcal{T}} & 0\\
\end{pmatrix}.
\end{equation}\newline

\section{Edge states in all 2D NH symmetry classes with intrinsic point gap topology}

In this section, we derive the edge signature of all intrinsically point-gapped topological NH symmetry classes in 2D.

\begin{figure}[b]
\centering
\includegraphics[width=0.5\textwidth]{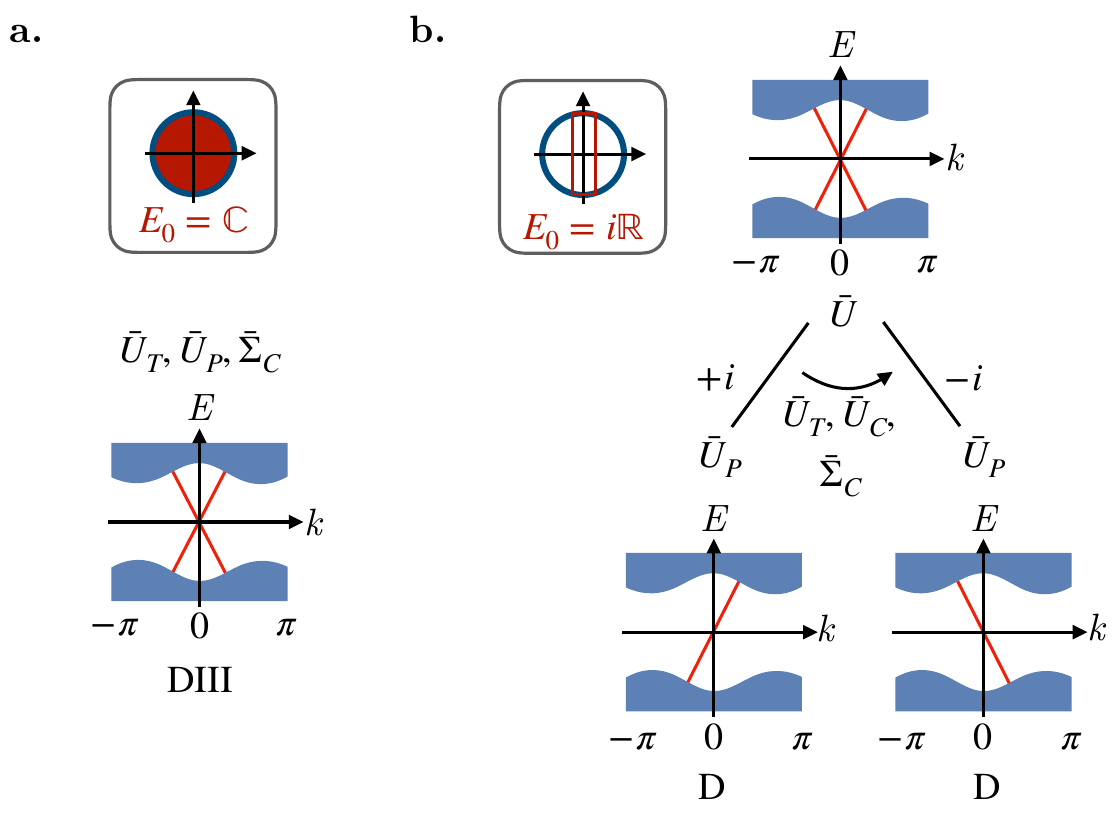}
\caption{\textbf{Classification of boundary physics for NH symmetry class AII$^\dagger$ and DIII$^\dagger$.} \textbf{a.} NH symmetry class AII$^\dagger$ corresponds to an EHH in Hermitian symmetry class DIII, whose non-trivial phase hosts a single Kramers pair at a time-reversal invariant momentum (TRIM). This results in an infernal point in the NH SIBC spectrum. \textbf{b.} For $E_0 \in i\mathbb{R}$, NH symmetry class DIII$^\dagger$ corresponds to having an EHH with two unitary subspaces in Hermitian symmetry class D. As outlined in Sec.~\ref{sec:sym-class:DIII_dag}, the corresponding non-trivial phase hosts a chiral boundary mode, which is paired to a counter-propagating partner by a TRS mapping between the unitary subspaces. The resulting EHH boundary spectrum, therefore, hosts a single Kramers pair. This results in an infernal point in the NH SIBC spectrum. Bulk states are depicted in blue and edge states are highlighted in red.\label{fig:sym-class:AIIDIII_dag}}
\end{figure}

\tocless\subsection{NH symmetry class AII$^\dagger$}{sec:sym-class:AII_dag}

NH symmetry class AII$^\dagger$ has a TRS$^\dagger$ squaring to minus one. As outlined in \ref{sec:sym-herm_ext}, the corresponding EHH at $E_0 =0$ introduces a chiral symmetry $\bar{\Sigma}_{\mathcal{C}}$, which combines with TRS to form a PHS with $\bar{U}_{\mathcal{P}}\bar{U}_{\mathcal{P}}^* = -\bar{U}_{\mathcal{T}}\bar{U}_{\mathcal{T}}^* = +1$. Consequently, the EHH is in Hermitian symmetry class DIII, which is $\ztwo$ classified for both 1D and 2D~\cite{Ryu2010}. The non-trivial 2D NH point-gapped phase thus corresponds to a single Kramers pair in the EHH boundary spectrum~\cite{Sato-PRL:2020}, pinned to zero energy and a TRIM (see Fig.~\ref{fig:sym-class:AIIDIII_dag}a). In the NH SIBC spectrum, these modes then correspond to edge-localized states at complex energy $E_0 = 0$ in the point gap, at a fixed TRIM.
NH symmetry class AII$^\dagger$ has a TRS$^\dagger$ squaring to minus one. As outlined in \ref{sec:sym-herm_ext}, the corresponding EHH at $E_0 =0$ introduces a chiral symmetry $\bar{\Sigma}_{\mathcal{C}}$, which combines with TRS to form a PHS with $\bar{U}_{\mathcal{P}}\bar{U}_{\mathcal{P}}^* = -\bar{U}_{\mathcal{T}}\bar{U}_{\mathcal{T}}^* = +1$. Consequently, the EHH is in Hermitian symmetry class DIII, which is $\ztwo$ classified for both 1D and 2D~\cite{Ryu2010}. The non-trivial 2D NH point-gapped phase thus corresponds to a single Kramers pair in the EHH boundary spectrum~\cite{Sato-PRL:2020}, pinned to zero energy and a TRIM (see Fig.~\ref{fig:sym-class:AIIDIII_dag}a). In the NH SIBC spectrum, these modes then correspond to edge-localized states at complex energy $E_0 = 0$ in the point gap, at a fixed TRIM.

\begin{figure*}[t!]
\centering
\includegraphics[width=1\textwidth,page=1]{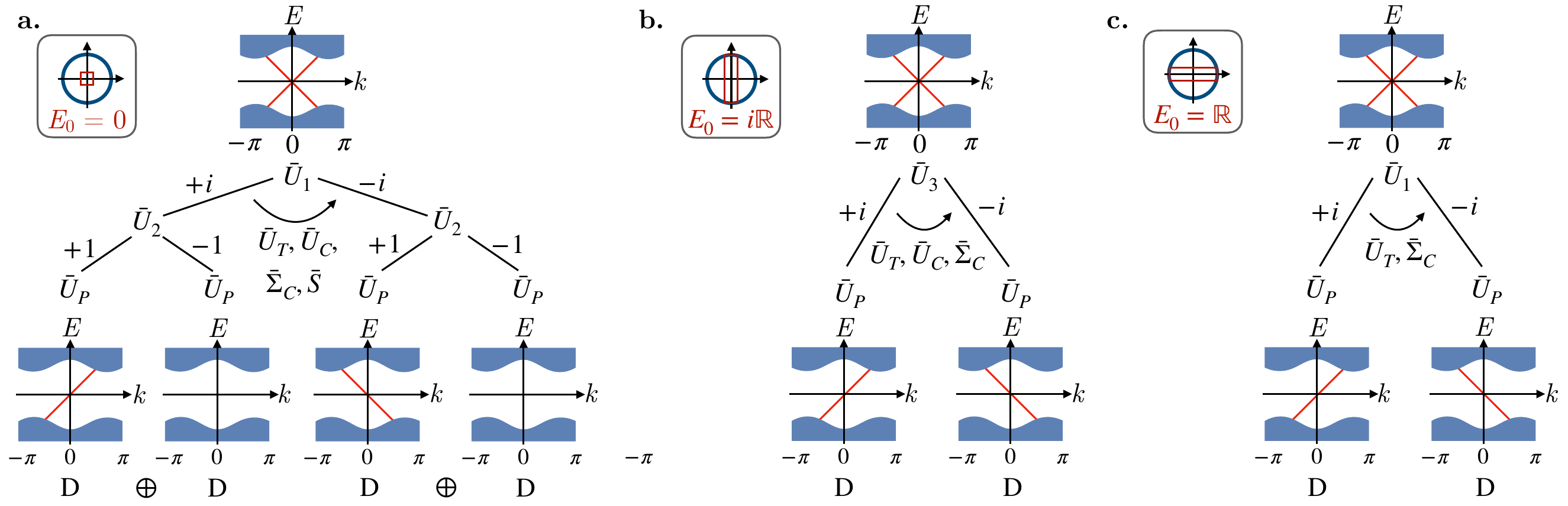}
\caption{\textbf{Classification of boundary physics for NH symmetry class BDI$^{S_{+-}}$.} \textbf{a.} The corresponding EHH at $E_0 =0$ obtains two unitary symmetries $\bar{U}_1, \bar{U}_2$, which allows to consider the respective unitary eigenspaces. The resulting classification outlined in Sec.~\ref{sec:sym-class:BDI_Spm} yields Hermitian symmetry class D$\oplus$D, with a Kramers pair at a TRIM. \textbf{b.} For $E_0 \in i\mathbb{R}$, the unitary symmetry $\bar{U}_1$ is replaced by $\bar{U}_3$, replicating the scenario of panel \textbf{a}. \textbf{c.} For $E_0 \in \mathbb{R}$, only the unitary symmetry $\bar{U}_1$ remains. As outlined in Sec.~\ref{sec:sym-class:BDI_Spm}, a chiral mode in one Hermitian symmetry class D subspace is mapped to its time-reversal symmetric partner in the other $\bar{U}_1$ subspace. Bulk states are depicted in blue and edge states are highlighted in red.\label{fig:sym-class:BDI_Spm}}
\end{figure*}

As the construction of the EHH can be repeated for every complex eigenvalue inside the point gap, the entire point gap of a corresponding NH system fills with boundary-localized modes under SIBC. Consequently, models in NH symmetry class AII$^\dagger$ show an infernal point.

\tocless\subsection{NH symmetry class DIII$^\dagger$}{sec:sym-class:DIII_dag}

NH symmetry class DIII$^\dagger$ possesses TRS, PHS and chiral symmetry with $(U_{\mathcal{T}}U_{\mathcal{T}}^*,U_{\mathcal{P}}U_{\mathcal{P}}^*) = (-1,1)$, where $U_{\mathcal{C}} = U_{\mathcal{T}}U_{\mathcal{P}}$. The corresponding EHH at $E_0 =0$ obtains TRS $\bar{U}_{\mathcal{T}}$, PHS $\bar{U}_{\mathcal{P}}$ and two chiral symmetries $\{\bar{U}_{\mathcal{C}},\bar{\Sigma}_{\mathcal{C}}\} = 0$. Both chiral symmetries can be combined to a unitary symmetry $\bar{U} = \bar{U}_{\mathcal{C}}\bar{\Sigma}_{\mathcal{C}}$ with $\bar{U}^2 = -1$.

Since $\bar{U}$ has an imaginary spectrum and $[\bar{U}_{\mathcal{T}},\bar{U}] = \{\bar{U}_{\mathcal{C}},\bar{U}\} = \{\bar{\Sigma}_{\mathcal{C}},\bar{U}\} = 0$, the eigenspaces of $\bar{U}$ are exchanged by anti-unitary TRS and the CSs. Since $\{\bar{U}_{\mathcal{P}},\bar{U}\} = 0$, we retain a PHS in each eigenspace, corresponding to Hermitian symmetry class D which has a $\z$ classification in 2D~\cite{Ryu2010}. After modding out line-gap phases, this is reduced to a $\ztwo$ classification~\cite{Sato-PRL:2020}. Consequently, point-gapped systems in NH symmetry class DIII$^\dagger$ are classified by a $\z$ valued Chern number $C_1$ which is restricted to values $C_1\in \{0,1\}$ by line-gapped phases~\cite{Sato-PRL:2020}.

\begin{figure}[b]
\centering
\includegraphics[width=0.38\textwidth,page=1]{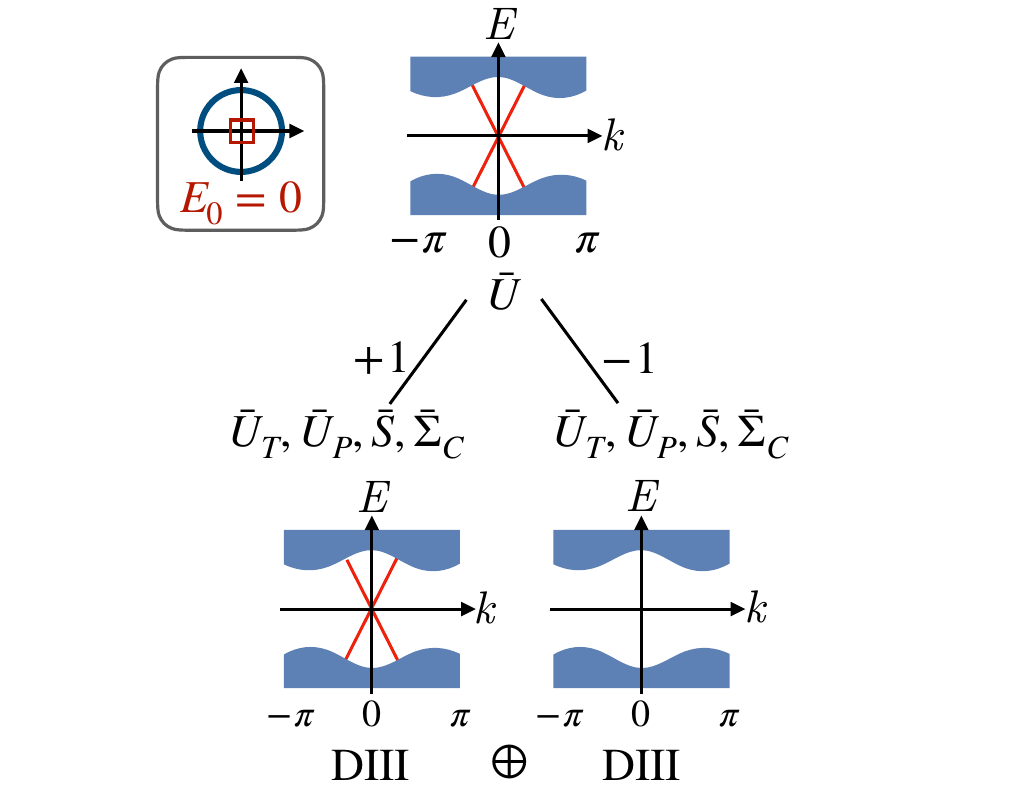}
\caption{\textbf{Classification of boundary physics for NH symmetry class D$^{S_{-}}$.} For $E_0=0$, NH symmetry class D$^{S_{-}}$ corresponds to having an EHH with two independent unitary subspaces in Hermitian symmetry class DIII. As outlined in Sec.~\ref{sec:sym-class:BDI_Spm}, the non-trivial phase shows one unitary subspace with a Kramers pair at a TRIM. This results in an infernal point in the NH SIBC spectrum. Bulk states are depicted in blue and edge states are highlighted in red.\label{fig:sym-class:D_Sm}}
\end{figure}

\begin{figure*}[t]
\centering
\includegraphics[width=1\textwidth,page=1]{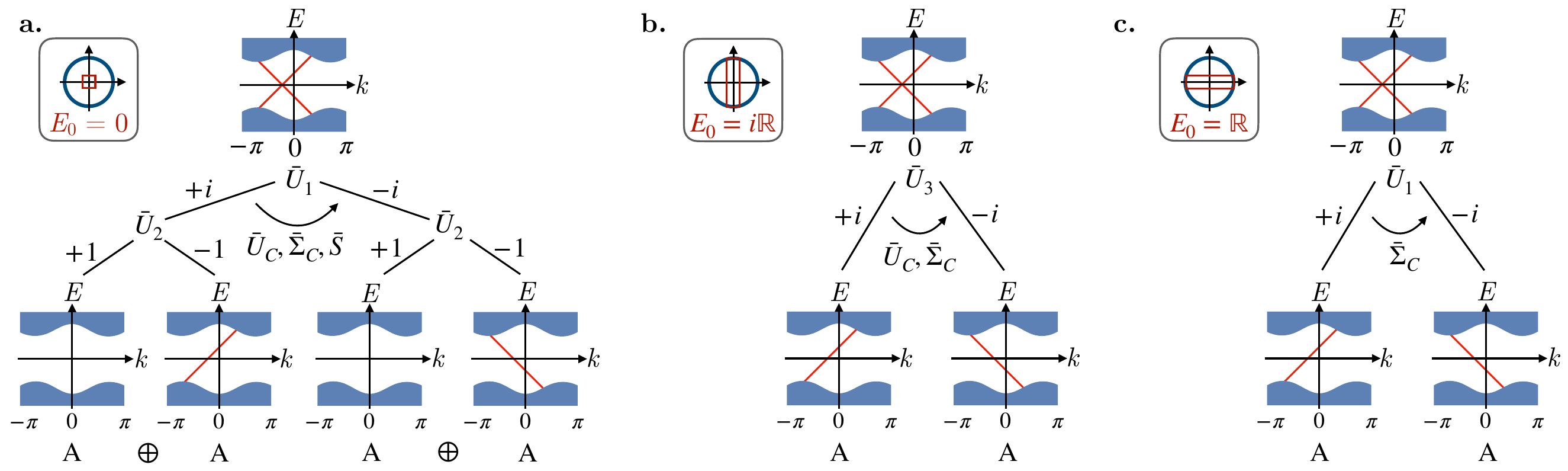}
\caption{\textbf{Classification of boundary physics for NH symmetry class AIII$^{S_{-}}$.} \textbf{a.} The corresponding EHH at $E_0 =0$ obtains two unitary symmetries $\bar{U}_1, \bar{U}_2$, which allows to consider the respective unitary eigenspaces. The resulting classification outlined in Sec.~\ref{sec:sym-class:AIIISm} yields Hermitian symmetry class A$\oplus$A, with two counter-propagating modes crossing at an arbitrary edge momentum. \textbf{b.} For $E_0 \in i\mathbb{R}$, the unitary symmetry $\bar{U}_1$ is replaced by $\bar{U}_3$, replicating the scenario of panel \textbf{a}. \textbf{c.} For $E_0 \in \mathbb{R}$, only the unitary symmetry $\bar{U}_1$ remains. As outlined in Sec.~\ref{sec:sym-class:AIIISm}, a chiral mode in one subspace is mapped to its chiral partner in the other $\bar{U}_1$ subspace. Bulk states are depicted in blue and edge states are highlighted in red.\label{fig:sym-class:AIIISm}}
\end{figure*}

In order to derive the signature of the nontrivial phase where $C_1=1$, we rely on the EHH for a given eigenvalue $E_0$ inside the point gap. For $E_0 = 0$, the EHH for NH symmetry class DIII$^\dagger$ enjoys the symmetries of Hermitian symmetry class D in each $\bar{U}$ eigenspace. Moreover, since Hermitian symmetry class D is $\z$-classified in 2D~\cite{Ryu2010}, the EHH associated with a nontrivial NH phase in symmetry class DIII$^\dagger$ must itself realize a nontrivial topological insulator phase in Hermitian symmetry class D. This non-trivial phase manifests as a chiral boundary mode, crossing zero energy at a TRIM and possibly further momenta. Since the eigenspaces of $\bar{U}$ are not independent, this mode is mapped through TRS and the CSs to a counter-propagating partner in the other $\bar{U}$ eigenspace. The full EHH thus hosts a Kramers pair crossing zero energy at least at a TRIM (see Fig.~\ref{fig:sym-class:AIIDIII_dag}b). In the NH SIBC spectrum, these modes then correspond to edge-localized states at complex eigenvalue $E_0 = 0$ in the point gap, at a fixed momentum.   

For $E_0 \in i\mathbb{R}$, we retain the unitary symmetry $\bar{U}$, TRS $\bar{U}_{\mathcal{T}}$, PHS $\bar{U}_{\mathcal{P}}$ and the two chiral symmetries $\bar{U}_{\mathcal{C}}, \bar{\Sigma}_{\mathcal{C}}$. This replicates the scenario at $E_0 = 0$ discussed above.

Away from $E_0 \in i\mathbb{R}$, we retain the TRS $\bar{U}_{\mathcal{T}}\bar{U}_{\mathcal{T}}^* = -1$ and CS $\bar{\Sigma}_{\mathcal{C}}$, which combine to form a PHS $\bar{U}_{\mathcal{P}} = \bar{\Sigma}_{\mathcal{C}} \bar{U}_{\mathcal{T}}$ with $\bar{U}_{\mathcal{P}}\bar{U}_{\mathcal{P}}^* = +1$. This corresponds to Hermitian symmetry class DIII, where due to TRS and CS, the Kramers pair remains protected at zero energy and a TRIM. Hence the entire point gap of a corresponding NH system fills with edge-localized modes at distinct momenta, the defining signature of an infernal point.\newline

\tocless\subsection{NH symmetry class BDI$^{S_{+-}}$}{sec:sym-class:BDI_Spm}

NH symmetry class BDI$^{S_{+-}}$ possesses TRS, PHS and CS with $(U_{\mathcal{T}}U_{\mathcal{T}}^*,U_{\mathcal{P}}U_{\mathcal{P}}^*) = (1,1)$, where $U_{\mathcal{C}} = U_{\mathcal{T}}U_{\mathcal{P}}$, as well as SLS $\mathcal{S}$. The corresponding EHH at $E_0 =0$ obtains TRS $\bar{U}_{\mathcal{T}}$, PHS $\bar{U}_{\mathcal{P}}$ and three CSs $\bar{U}_{\mathcal{C}}, \bar{\Sigma}_{\mathcal{C}}, \bar{\mathcal{S}}$. The CSs can be combined to two commuting unitary symmetries $\bar{U}_1 = \bar{U}_{\mathcal{C}}\bar{\mathcal{S}}$ with $\bar{U}_1^2 = -1$ and $\bar{U}_2 = \bar{\mathcal{S}} \bar{\Sigma}_{\mathcal{C}}$ with $\bar{U}_2^2 = +1$.

Since $\bar{U}_1$ has an imaginary spectrum and $[\bar{U}_{\mathcal{T}},\bar{U}_1] = \{\bar{U}_{\mathcal{P}},\bar{U}_1\} = \{\bar{U}_{\mathcal{C}},\bar{U}_1\} = \{\bar{\Sigma}_{\mathcal{C}},\bar{U}_1\} = \{\bar{\mathcal{S}},\bar{U}_1\}$, the eigenspaces of $\bar{U}_1$ are not independent and individually enjoy $\bar{U}_{\mathcal{P}}$ and $\bar{U}_2$ symmetry. Moreover, we have $[\bar{U}_{\mathcal{T}},\bar{U}_2] = [\bar{U}_{\mathcal{P}},\bar{U}_2] = [\bar{\mathcal{S}},\bar{U}_2] = [\bar{U}_{\mathcal{C}},\bar{U}_2] = [\bar{\Sigma}_{\mathcal{C}},\bar{U}_2] =0$, so that the $\bar{U}_2$ eigenspaces are independent and individually preserve $\bar{U}_{\mathcal{P}}$ symmetry. They therefore lie in Hermitian symmetry class D and yield a $\z \oplus \z$ classification in 2D~\cite{Ryu2010} that is reduced to $\ztwo$ by line gap phases~\cite{Sato-PRL:2020}. The non-trivial point gap phase is the one where only one $\bar{U}_2$ subspace is non-trivial, hosting a chiral boundary mode, crossing zero energy at a TRIM and possibly further momenta. Since the eigenspaces of $\bar{U}_1$ are not independent, this mode is mapped through TRS and the CSs to a counter-propagating partner in the other $\bar{U}_1$ eigenspace. The full EHH thus hosts a Kramers pair crossing zero energy at least at a TRIM (see Fig.~\ref{fig:sym-class:BDI_Spm}a). In the NH SIBC spectrum, these modes then correspond to edge-localized states at complex eigenvalue $E_0 = 0$ in the point gap, at a fixed momentum.   

For $E_0 \in i\mathbb{R}$, we retain the two chiral symmetries $\bar{U}_{\mathcal{C}}, \bar{\Sigma}_{\mathcal{C}}$. Additionally, we can form the TRS $\bar{U}_{\mathcal{T}} = \bar{\mathcal{S}} \bar{U}_{\mathcal{P}}$ [Eq.~\eqref{eq:altTRS}] with $\bar{U}_{\mathcal{T}}\bar{U}_{\mathcal{T}}^* = -1$ and PHS $\bar{U}_{\mathcal{P}} = \bar{\Sigma}_{\mathcal{C}} \bar{U}_{\mathcal{T}}$ with $\bar{U}_{\mathcal{P}}\bar{U}_{\mathcal{P}}^* = +1$ [Eq.~\eqref{eq:altPHS}]. This allows to define an additional unitary symmetry $\bar{U}_3 = \bar{U}_{\mathcal{C}} \bar{\Sigma}_{\mathcal{C}}$ with $\bar{U}_3^2 = -1$, which anticommutes with $ \bar{U}_{\mathcal{P}}, \bar{U}_{\mathcal{C}}, \bar{\Sigma}_{\mathcal{C}}$ but commutes with $\bar{U}_{\mathcal{T}}$. This corresponds to Hermitian symmetry class D, with a chiral mode per unitary subspace, crossing zero energy at a TRIM (see Fig.~\ref{fig:sym-class:BDI_Spm}b).

For $E_0 \in \mathbb{R}$, we are only left with $\bar{U}_1$, whose sectors are mapped onto each other by $\bar{\Sigma}_{\mathcal{C}}, \bar{U}_{\mathcal{T}}$ [Eq.~\eqref{eq:altTRS}] but retain $\bar{U}_{\mathcal{P}}$ [Eq.~\eqref{eq:altPHS}]. This corresponds to Hermitian symmetry class D, with a chiral mode per unitary subspace, crossing zero energy at a TRIM (see Fig.~\ref{fig:sym-class:BDI_Spm}c). 

For arbitrary $E_0 \in \mathbb{C}$, we have $\bar{U}_{\mathcal{T}}$ [Eq.~\eqref{eq:altTRS}], $\bar{U}_{\mathcal{P}}$ [Eq.~\eqref{eq:altPHS}], $\bar{\Sigma}_{\mathcal{C}}$ left, and hence Hermitian symmetry class DIII. Due to TRS and CS, the Kramers pair remains protected at zero energy and a TRIM. Hence the entire point gap of a corresponding NH system fills with edge-localized modes at distinct momenta, the defining signature of an infernal point.\newline

\begin{figure*}[t!]
\centering
\includegraphics[width=1\textwidth,page=1]{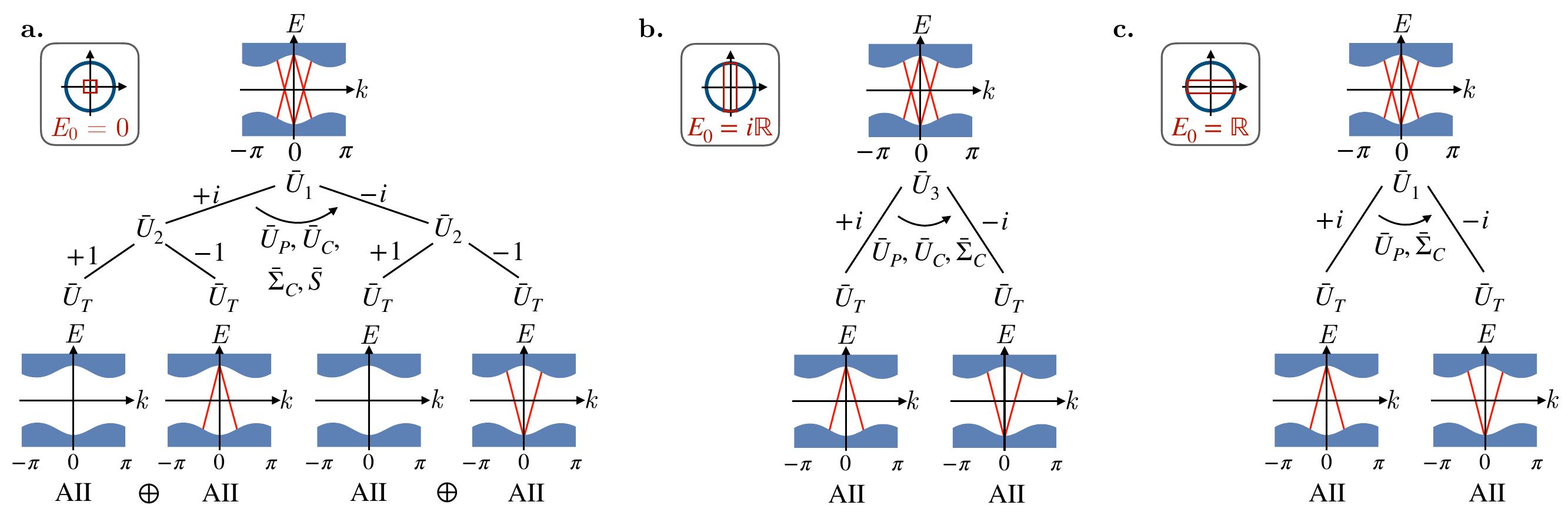}
\caption{\textbf{Classification of boundary physics for NH symmetry class DIII$^{S_{+-}}$.} \textbf{a.} The corresponding EHH at $E_0 =0$ obtains two unitary symmetries $\bar{U}_1, \bar{U}_2$, which allows to consider the respective unitary eigenspaces. The resulting classification outlined in Sec.~\ref{sec:sym-class:DIIISpm} yields Hermitian symmetry class AII$\oplus$AII, with two Kramers pairs crossing at symmetric momenta around a TRIM. \textbf{b.} For $E_0 \in i\mathbb{R}$, the unitary symmetry $\bar{U}_1$ is replaced by $\bar{U}_3$, replicating the scenario of panel \textbf{a}. \textbf{c.} For $E_0 \in \mathbb{R}$, only the unitary symmetry $\bar{U}_1$ remains. As outlined in Sec.~\ref{sec:sym-class:DIIISpm}, a Kramers pair in one subspace is mapped to its particle-hole symmetric partner in the other $\bar{U}_1$ subspace. Bulk states are depicted in blue and edge states are highlighted in red.\label{fig:sym-class:DIIISpm}}
\end{figure*}

\tocless\subsection{NH symmetry class D$^{S_{-}}$}{sec:sym-class:D_Sm}

Models in NH symmetry class D$^{S_{-}}$ possess PHS $U_{\mathcal{P}}$ ($U_{\mathcal{P}}U_{\mathcal{P}}^* = +1$) and SLS $\mathcal{S}$, with $\{U_{\mathcal{P}},\mathcal{S}\} = 0$. The corresponding EHH at $E_0 =0$ introduces a CS $\bar{\Sigma}_{\mathcal{C}}$, which combines with SLS to form a unitary symmetry $\bar{U} = \bar{\mathcal{S}}\bar{\Sigma}_{\mathcal{C}}$ with $\bar{U}^2 = +1$. The unitary $\bar{U}$ satisifies $[\bar{U}_{\mathcal{P}},\bar{U}]=[\bar{\mathcal{S}},\bar{U}]=[\bar{\Sigma}_{\mathcal{C}},\bar{U}]=0$. We may furthermore define a TRS $\bar{U}_{\mathcal{T}} = \bar{\mathcal{S}} \bar{U}_{\mathcal{P}}$ which satisfies $\bar{U}_{\mathcal{T}}\bar{U}_{\mathcal{T}}^* = -1$. Hence we obtain Hermitian symmetry class DIII in each $\bar{U}$ subspace, giving a $\ztwo \oplus \ztwo$ classification in 2D~\cite{Ryu2010}. By modding out line-gap phases, this is reduced to a $\ztwo$ classification where the non-trivial element corresponds to having only a single $\bar{U}$ subspace being non-trivial~\cite{Sato-PRL:2020}. This non-trivial phase manifests as a Kramers pair on the boundary, crossing zero energy at a TRIM (see Fig.~\ref{fig:sym-class:D_Sm}b). In the NH SIBC spectrum, these modes then correspond to edge-localized states at complex eigenvalue $E_0 = 0$ in the point gap, at a fixed momentum.

For $E_0 \in i\mathbb{R}$, we retain TRS $\bar{U}_{\mathcal{T}}$ [Eq.~\eqref{eq:altTRS}], PHS $\bar{U}_{\mathcal{P}}$ [Eq.~\eqref{eq:altPHS}] and the CS $\bar{\Sigma}_{\mathcal{C}}$. This corresponds to Hermitian symmetry class DIII, where due to TRS and CS, the Kramers pair remains protected at zero energy and a TRIM.

Away from $E_0 \in i\mathbb{R}$, we retain the TRS $\bar{U}_{\mathcal{T}}\bar{U}_{\mathcal{T}}^* = -1$ defined above as well as CS $\bar{\Sigma}_{\mathcal{C}}$, which combine to form a PHS $\bar{U}_{\mathcal{P}} = \bar{\Sigma}_{\mathcal{C}} \bar{U}_{\mathcal{T}}$ with $\bar{U}_{\mathcal{P}}\bar{U}_{\mathcal{P}}^* = +1$. This corresponds to Hermitian symmetry class DIII, where due to TRS and CS, the Kramers pair remains protected at zero energy and a TRIM. Hence the entire point gap of a corresponding NH system fills with edge-localized modes at distinct momenta, the defining signature of an infernal point.\newline

\tocless\subsection{NH symmetry class AIII$^{S_{-}}$}{sec:sym-class:AIIISm}

Models in NH symmetry class AIII$^{S_{-}}$ possess CS $U_{\mathcal{C}}$ and SLS $\mathcal{S}$, with $\{U_{\mathcal{C}},\mathcal{S}\} = 0$. The corresponding EHH at $E_0 =0$ obtains three CSs $\bar{U}_{\mathcal{C}}, \bar{\Sigma}_{\mathcal{C}}, \bar{\mathcal{S}}$, which can be combined to two unitary symmetries $\bar{U}_1 = \bar{U}_{\mathcal{C}}\bar{\mathcal{S}}$ with $\bar{U}_1^2 = -1$ and $\bar{U}_2 = \bar{\mathcal{S}} \bar{\Sigma}_{\mathcal{C}}$ with $\bar{U}_2^2 = +1$. Due to the anti-commutation of $\bar{U}_1$ with the CSs, the eigenspaces of $\bar{U}_1$ are exchanged by $\bar{U}_{\mathcal{C}}, \bar{\Sigma}_{\mathcal{C}}, \bar{\mathcal{S}}$. Since $[\bar{U}_1,\bar{U}_2] = 0$, we retain $\bar{U}_2$ in each subspace. However, each eigenspace of $\bar{U}_2$ has no symmetry left, corresponding to Hermitian symmetry class A$\oplus$A, which has a $\z \oplus \z$ classification in 2D~\cite{Ryu2010}. After modding out line-gap phases, this is reduced to a $\ztwo$ classification, with only one subspace being non-trivial~\cite{Sato-PRL:2020}. The chiral edge mode of a non-trivial model in Hermitian symmetry class A can cross zero energy at an arbitrary edge momentum $k$, which is in general not sampled over in the discrete edge Brillouin zone associated with any finite system size. Hence, there are no exact zero-energy states in the EHH boundary spectrum, and we do not expect to observe an infernal point. The resulting edge state disperses as a function of $k$, forming a NH edge state. Note however that the action of $\bar{U}_{\mathcal{C}}, \bar{\Sigma}_{\mathcal{C}}, \bar{\mathcal{S}}$ pairs the chiral mode in one $\bar{U}_1$ subspace with a counter-propagating partner in the other $\bar{U}_1$ subspace, as seen in Fig.~\ref{fig:sym-class:AIIISm}a.

For $E_0 \in i\mathbb{R}$, we retain $\bar{U}_{\mathcal{C}}, \bar{\Sigma}_{\mathcal{C}}$. This allows to form an additional unitary symmetry $\bar{U}_3 = \bar{U}_{\mathcal{C}} \bar{\Sigma}_{\mathcal{C}}$ with $\bar{U}_3^2 = -1$, which anticommutes with $\bar{U}_{\mathcal{C}}, \bar{\Sigma}_{\mathcal{C}}$. Each $\bar{U}_3$ subspace thus has no symmetry left and remains in Hermitian symmetry class A with one chiral mode, but is mapped by the chiral symmetries. The crossing, counter-propagating modes at an arbitrary edge momentum therefore persist, as shown in Fig.~\ref{fig:sym-class:AIIISm}b.   

For $E_0 \in \mathbb{R}$, we are only left with $\bar{U}_1$, whose sectors are mapped onto each other by $\bar{\Sigma}_{\mathcal{C}}$. This corresponds to Hermitian symmetry class A. Each subspace of $\bar{U}_1$, therefore, still retains a chiral mode (see Fig.~\ref{fig:sym-class:AIIISm}c).

For $E_0 \in \mathbb{C}$, only the CS $\bar{\Sigma}_{\mathcal{C}}$ is present. A crossing of counter-propagating modes can be gapped while respecting CS, removing in-gap states in the EHH boundary spectrum. As a result, the NH edge state disperses as a function of $k$, but only along the real and imaginary axis. This dispersion forms a single exceptional point on the edge.\newline

\begin{figure*}[t]
\centering
\includegraphics[width=1\textwidth,page=1]{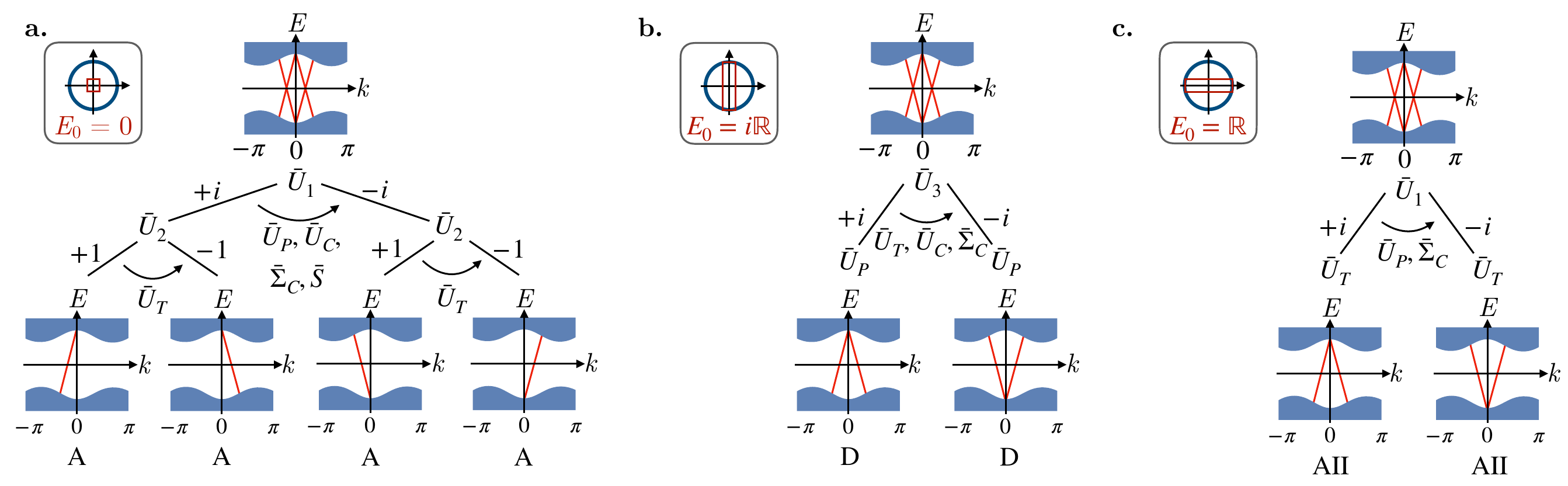}
\caption{\textbf{Classification of boundary physics for NH symmetry class CII$^{S_{-+}}$.} \textbf{a.} The corresponding EHH at $E_0 =0$ obtains two unitary symmetries $\bar{U}_1, \bar{U}_2$, which allows to consider the respective unitary eigenspaces. The resulting classification outlined in Sec.~\ref{sec:sym-class:CII_Smp} yields Hermitian symmetry class A, with two Kramers pairs crossing at symmetric momenta around a TRIM. \textbf{b.} For $E_0 \in i\mathbb{R}$, the unitary symmetry $\bar{U}_1$ is replaced by $\bar{U}_3$, replicating the scenario of panel \textbf{a}. \textbf{c.} For $E_0 \in \mathbb{R}$, only the unitary symmetry $\bar{U}_1$ remains. As outlined in Sec.~\ref{sec:sym-class:CII_Smp}, a Kramers pair in one subspace is mapped to its particle-hole symmetric partner in the other $\bar{U}_1$ subspace. Bulk states are depicted in blue and edge states are highlighted in red.\label{fig:sym-class:CII_Smp}}
\end{figure*}

\tocless\subsection{NH symmetry class DIII$^{S_{+-}}$}{sec:sym-class:DIIISpm}

NH symmetry class DIII$^{S_{+-}}$ possesses TRS, PHS and CS with $(U_{\mathcal{T}}U_{\mathcal{T}}^*,U_{\mathcal{P}}U_{\mathcal{P}}^*) = (-1,1)$, where $U_{\mathcal{C}} = U_{\mathcal{T}}U_{\mathcal{P}}$, as well as SLS $\mathcal{S}$. The corresponding EHH at $E_0 =0$ obtains TRS $\bar{U}_{\mathcal{T}}$, PHS $\bar{U}_{\mathcal{P}}$ and three CSs $\bar{U}_{\mathcal{C}}, \bar{\Sigma}_{\mathcal{C}}, \bar{\mathcal{S}}$. The CSs can be combined to two commuting unitary symmetries $\bar{U}_1 = \bar{U}_{\mathcal{C}}\bar{\mathcal{S}}$ with $\bar{U}_1^2 = -1$ and $\bar{U}_2 = \bar{\mathcal{S}} \bar{\Sigma}_{\mathcal{C}}$ with $\bar{U}_2^2 = +1$.

Since $\bar{U}_1$ has an imaginary spectrum and $\{\bar{U}_{\mathcal{T}},\bar{U}_1\} = [\bar{U}_{\mathcal{P}},\bar{U}_1] = \{\bar{U}_{\mathcal{C}},\bar{U}_1\} = \{\bar{\Sigma}_{\mathcal{C}},\bar{U}_1\} = \{\bar{\mathcal{S}},\bar{U}_1\}$, the eigenspaces of $\bar{U}_1$ are not independent and individually enjoy $\bar{U}_{\mathcal{T}}$ and $\bar{U}_2$ symmetry. Moreover, we have $[\bar{U}_{\mathcal{T}},\bar{U}_2] = [\bar{U}_{\mathcal{P}},\bar{U}_2] = [\bar{\mathcal{S}},\bar{U}_2] = [\bar{U}_{\mathcal{C}},\bar{U}_2] = [\bar{\Sigma}_{\mathcal{C}},\bar{U}_2] =0$, so that the $\bar{U}_2$ eigenspaces are independent and individually preserve $\bar{U}_{\mathcal{T}}$ symmetry. They therefore lie in Hermitian symmetry class AII and yield a $\ztwo \oplus \ztwo$ classification in 2D~\cite{Ryu2010} that is reduced to $\ztwo$ by line-gap phases~\cite{Sato-PRL:2020}. The non-trivial element corresponds to having only a single $\bar{U}_2$ subspace non-trivial. The helical edge modes of a non-trivial model in Hermitian symmetry class AII can cross zero energy at an arbitrary edge momentum $k$, which is in general not sampled over in the discrete edge Brillouin zone associated with any finite system size. Hence, there are no exact zero-energy states in the EHH boundary spectrum, and we do not expect to observe an infernal point. The resulting edge state disperses as a function of $k$, forming a NH edge state. Due to the presence of $\bar{U}_{\mathcal{P}}$, which maps between $\bar{U}_1$ subspaces, the entire EHH boundary spectrum hosts two Kramers pairs at symmetric momenta around a TRIM (see Fig.~\ref{fig:sym-class:DIIISpm}a).

For $E_0 \in i\mathbb{R}$, we retain $\bar{U}_{\mathcal{T}}$ [Eq.~\eqref{eq:altTRS}], $\bar{U}_{\mathcal{P}}$ [Eq.~\eqref{eq:altPHS}], $\bar{U}_{\mathcal{C}}, \bar{\Sigma}_{\mathcal{C}}$. This allows to form an additional unitary symmetry $\bar{U}_3 = \bar{U}_{\mathcal{C}} \bar{\Sigma}_{\mathcal{C}}$ with $\bar{U}_3^2 = -1$, which anticommutes with $ \bar{U}_{\mathcal{T}}, \bar{U}_{\mathcal{C}}, \bar{\Sigma}_{\mathcal{C}}$ but commutes with $\bar{U}_{\mathcal{P}}$. Each $\bar{U}_3$ subspace thus possesses the TRS $\bar{U}_{\mathcal{T}}$. The eigensectors of $\bar{U}_3$ therefore are in Hermitian symmetry class AII. The Kramers pair in one unitary subspace is mapped onto one in the other subspace by the action of $\bar{U}_{\mathcal{P}}, \bar{U}_{\mathcal{C}}, \bar{\Sigma}_{\mathcal{C}}$. The two Kramers pairs at symmetric momenta around a TRIM therefore persist, as shown in Fig.~\ref{fig:sym-class:AIIISm}b.

For $E_0 \in \mathbb{R}$, we are only left with $\bar{U}_1$, whose sectors are mapped onto each other by $\bar{\Sigma}_{\mathcal{C}}, \bar{U}_{\mathcal{P}}$ [Eq.~\eqref{eq:altPHS}] but retain $\bar{U}_{\mathcal{T}}$ [Eq.~\eqref{eq:altTRS}]. This corresponds to Hermitian symmetry class AII, with a single Kramers pair per unitary subspace. 

For arbitrary $E_0 \in \mathbb{C}$, we have $\bar{U}_{\mathcal{T}}$ [Eq.~\eqref{eq:altTRS}], $\bar{U}_{\mathcal{P}}$ [Eq.~\eqref{eq:altPHS}], $\bar{\Sigma}_{\mathcal{C}}$ left, and hence Hermitian symmetry class DIII. The two Kramers pairs are no longer protected, removing in-gap states in the EHH boundary spectrum. As a result, the NH edge state disperses as a function of $k$, but only along the real and imaginary axis. This dispersion forms a pair of exceptional points on the edge, represented in the EHH by two Kramers pairs.\newline

\tocless\subsection{NH symmetry class CII$^{S_{-+}}$}{sec:sym-class:CII_Smp}

NH symmetry class CII$^{S_{-+}}$ possesses TRS, PHS and CS with $(U_{\mathcal{T}}U_{\mathcal{T}}^*,U_{\mathcal{P}}U_{\mathcal{P}}^*) = (-1,-1)$, where $U_{\mathcal{C}} = U_{\mathcal{T}}U_{\mathcal{P}}$, as well as SLS $\mathcal{S}$. The corresponding EHH at $E_0 =0$ obtains TRS $\bar{U}_{\mathcal{T}}$, PHS $\bar{U}_{\mathcal{P}}$ and three CSs $\bar{U}_{\mathcal{C}}, \bar{\Sigma}_{\mathcal{C}}, \bar{\mathcal{S}}$. The CSs can be combined to two commuting unitary symmetries $\bar{U}_1 = \bar{U}_{\mathcal{C}}\bar{\mathcal{S}}$ with $\bar{U}_1^2 = -1$ and $\bar{U}_2 = \bar{\mathcal{S}} \bar{\Sigma}_{\mathcal{C}}$ with $\bar{U}_2^2 = +1$.

\begin{figure*}[t]
\centering
\includegraphics[width=1\textwidth,page=1]{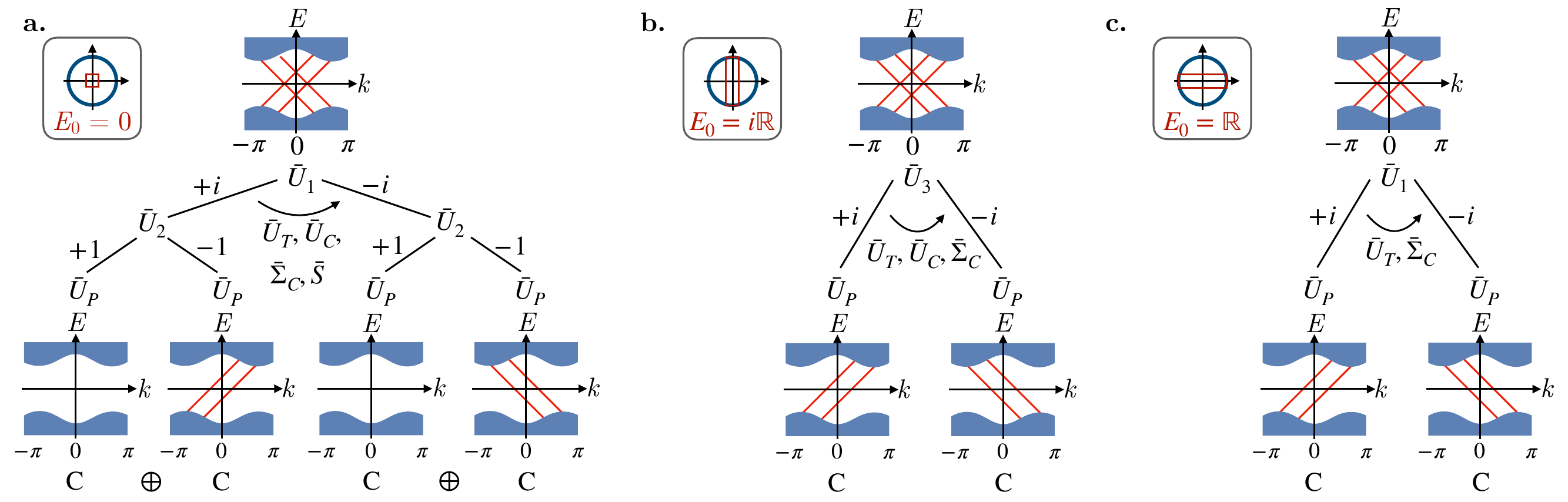}
\caption{\textbf{Classification of boundary physics for NH symmetry class CII$^{S_{+-}}$.} \textbf{a.} The corresponding EHH at $E_0 =0$ obtains two unitary symmetries $\bar{U}_1, \bar{U}_2$, which allows to consider the respective unitary eigenspaces. The resulting classification outlined in Sec.~\ref{sec:sym-class:CII_Spm} yields Hermitian symmetry class C$\oplus$C, with two Kramers pairs crossing at symmetric momenta around a TRIM. \textbf{b.} For $E_0 \in i\mathbb{R}$, the unitary symmetry $\bar{U}_1$ is replaced by $\bar{U}_3$, replicating the scenario of panel \textbf{a}. \textbf{c.} For $E_0 \in \mathbb{R}$, only the unitary symmetry $\bar{U}_1$ remains. As outlined in Sec.~\ref{sec:sym-class:CII_Spm}, a pair of chiral modes in one Hermitian symmetry class C subspace is mapped to its time-reversal symmetric partner in the other $\bar{U}_1$ subspace. Bulk states are depicted in blue and edge states are highlighted in red.\label{fig:sym-class:CII_Spm}}
\end{figure*}

Since $\bar{U}_1$ has an imaginary spectrum and $[\bar{U}_{\mathcal{P}},\bar{U}_1] = \{\bar{U}_{\mathcal{T}},\bar{U}_1\} = \{\bar{U}_{\mathcal{C}},\bar{U}_1\} = \{\bar{\Sigma}_{\mathcal{C}},\bar{U}_1\} = \{\bar{\mathcal{S}},\bar{U}_1\}$, the eigenspaces of $\bar{U}_1$ are not independent and individually enjoy $\bar{U}_{\mathcal{T}}$ and $\bar{U}_2$ symmetry. Moreover, we have $\{\bar{U}_{\mathcal{T}},\bar{U}_2\} = \{\bar{U}_{\mathcal{P}},\bar{U}_2\} = [\bar{\mathcal{S}},\bar{U}_2] = [\bar{U}_{\mathcal{C}},\bar{U}_2] = [\bar{\Sigma}_{\mathcal{C}},\bar{U}_2] =0$, so that the $\bar{U}_2$ eigenspaces are exchanged by $\bar{U}_{\mathcal{T}}$ symmetry. This leaves us with Hermitian symmetry class A, which has a $\z$ classification in 2D~\cite{Ryu2010} that is reduced to $\ztwo$ by line-gap phases~\cite{Sato-PRL:2020}. The non-trivial element corresponds to having a single chiral mode in one $\bar{U}_2$ subspace. The chiral edge mode of a non-trivial model in Hermitian symmetry class A can cross zero energy at an arbitrary edge momentum $k$, which is in general not sampled over in the discrete edge Brillouin zone associated with any finite system size. Hence, there are no exact zero-energy states in the EHH boundary spectrum, and we do not expect to observe an infernal point. The resulting edge state disperses as a function of $k$, forming a NH edge state. Due to the presence of $\bar{U}_{\mathcal{T}}$, which maps between $\bar{U}_2$ subspaces, and $\bar{U}_{\mathcal{P}}$, which maps between $\bar{U}_1$ subspaces, the entire EHH boundary spectrum hosts two Kramers pairs at symmetric momenta around a TRIM (see Fig.~\ref{fig:sym-class:CII_Smp}a).

For $E_0 \in i\mathbb{R}$, we retain $\bar{U}_{\mathcal{T}}$ [Eq.~\eqref{eq:altTRS}], $\bar{U}_{\mathcal{P}}$ [Eq.~\eqref{eq:altPHS}], $\bar{U}_{\mathcal{C}}, \bar{\Sigma}_{\mathcal{C}}$. This allows to form an additional unitary symmetry $\bar{U}_3 = \bar{U}_{\mathcal{C}} \bar{\Sigma}_{\mathcal{C}}$ with $\bar{U}_3^2 = -1$, which anticommutes with $ \bar{U}_{\mathcal{P}}, \bar{U}_{\mathcal{C}}, \bar{\Sigma}_{\mathcal{C}}$ but commutes with $\bar{U}_{\mathcal{T}}$. Each $\bar{U}_3$ subspace thus possesses the PHS $\bar{U}_{\mathcal{P}}$, resulting in Hermitian symmetry class D with two chiral modes. The eigensectors of $\bar{U}_3$ are exchanged by $ \bar{U}_{\mathcal{T}}, \bar{U}_{\mathcal{C}}, \bar{\Sigma}_{\mathcal{C}}$, forming two Kramers pairs at symmetric momenta around a TRIM, as shown in Fig.~\ref{fig:sym-class:CII_Smp}b.

For $E_0 \in \mathbb{R}$, we are only left with $\bar{U}_1$, whose sectors are mapped onto each other by $\bar{\Sigma}_{\mathcal{C}}, \bar{U}_{\mathcal{P}}$  [Eq.~\eqref{eq:altPHS}] but retain $\bar{U}_{\mathcal{T}}$ [Eq.~\eqref{eq:altTRS}]. This corresponds to Hermitian symmetry class AII, with helical modes in one subspace, mapped to their particle-hole symmetric partners in the other $\bar{U}_1$ subspace. The entire EHH boundary spectrum thus still contains two Kramers pairs at symmetric momenta around a TRIM (see Fig.~\ref{fig:sym-class:CII_Smp}c).

For arbitrary $E_0 \in \mathbb{C}$, we have $\bar{U}_{\mathcal{T}}$ [Eq.~\eqref{eq:altTRS}], $\bar{U}_{\mathcal{P}}$ [Eq.~\eqref{eq:altPHS}], $\bar{\Sigma}_{\mathcal{C}}$ left, and hence Hermitian symmetry class DIII. The two Kramers pairs are no longer protected, removing in-gap states in the EHH boundary spectrum. As a result, the NH edge state disperses as a function of $k$, but only along the real and imaginary axis. This dispersion forms a pair of exceptional points on the edge, represented in the EHH by two Kramers pairs.\newline

\begin{figure*}[t]
\centering
\includegraphics[width=1\textwidth,page=1]{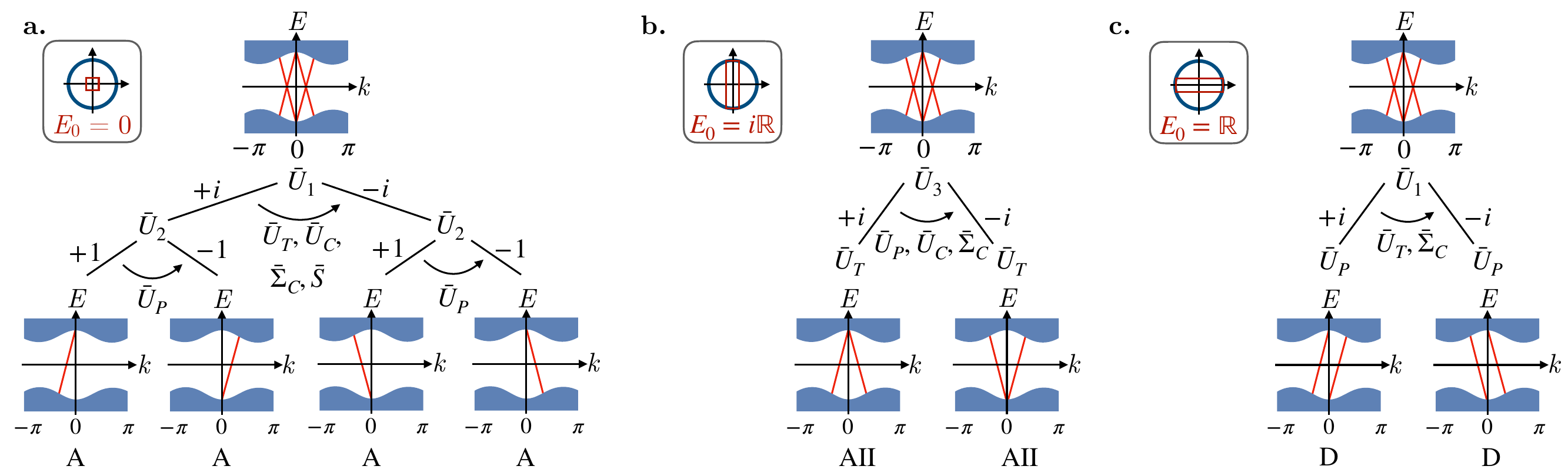}
\caption{\textbf{Classification of boundary physics for NH symmetry class CI$^{S_{-+}}$.} \textbf{a.} The corresponding EHH at $E_0 =0$ obtains two unitary symmetries $\bar{U}_1, \bar{U}_2$, which allows to consider the respective unitary eigenspaces. The resulting classification outlined in Sec.~\ref{sec:sym-class:CI_Smp} yields Hermitian symmetry class A, with two Kramers pairs crossing at symmetric momenta around a TRIM. \textbf{b.} For $E_0 \in i\mathbb{R}$, the unitary symmetry $\bar{U}_1$ is replaced by $\bar{U}_3$, replicating the scenario of panel \textbf{a}. \textbf{c.} For $E_0 \in \mathbb{R}$, only the unitary symmetry $\bar{U}_1$ remains. As outlined in Sec.~\ref{sec:sym-class:CI_Smp}, a pair of chiral modes in one subspace is mapped to its time-reversal symmetric partner in the other $\bar{U}_1$ subspace. Bulk states are depicted in blue and edge states are highlighted in red.\label{fig:sym-class:CI_Smp}}
\end{figure*}

\tocless\subsection{NH symmetry class CII$^{S_{+-}}$}{sec:sym-class:CII_Spm}

The NH symmetry class CII$^{S_{+-}}$ possesses TRS, PHS and CS with $(U_{\mathcal{T}}U_{\mathcal{T}}^*,U_{\mathcal{P}}U_{\mathcal{P}}^*) = (-1,-1)$, where $U_{\mathcal{C}} = U_{\mathcal{T}}U_{\mathcal{P}}$, as well as SLS $\mathcal{S}$. The corresponding EHH at $E_0 =0$ obtains TRS $\bar{U}_{\mathcal{T}}$, PHS $\bar{U}_{\mathcal{P}}$ and three CSs $\bar{U}_{\mathcal{C}}, \bar{\Sigma}_{\mathcal{C}}, \bar{\mathcal{S}}$. The CSs can be combined to two commuting unitary symmetries $\bar{U}_1 = \bar{U}_{\mathcal{C}}\bar{\mathcal{S}}$ with $\bar{U}_1^2 = -1$ and $\bar{U}_2 = \bar{\mathcal{S}} \bar{\Sigma}_{\mathcal{C}}$ with $\bar{U}_2^2 = +1$.

Since $\bar{U}_1$ has an imaginary spectrum and $\{\bar{U}_{\mathcal{P}},\bar{U}_1\} = [\bar{U}_{\mathcal{T}},\bar{U}_1] = \{\bar{U}_{\mathcal{C}},\bar{U}_1\} = \{\bar{\Sigma}_{\mathcal{C}},\bar{U}_1\} = \{\bar{\mathcal{S}},\bar{U}_1\}$, the eigenspaces of $\bar{U}_1$ are not independent and individually enjoy $\bar{U}_{\mathcal{P}}$ and $\bar{U}_2$ symmetry. Moreover, we have $[\bar{U}_{\mathcal{T}},\bar{U}_2] = [\bar{U}_{\mathcal{P}},\bar{U}_2] = [\bar{\mathcal{S}},\bar{U}_2] = [\bar{U}_{\mathcal{C}},\bar{U}_2] = [\bar{\Sigma}_{\mathcal{C}},\bar{U}_2] =0$, so that the $\bar{U}_2$ eigenspaces are independent and individually preserve $\bar{U}_{\mathcal{P}}$ symmetry. They therefore lie in Hermitian symmetry class C and yield a $2\z \oplus 2\z$ classification in 2D~\cite{Ryu2010} that is reduced to $\ztwo$ by line-gap phases~\cite{Sato-PRL:2020}. The non-trivial element corresponds to having only a single $\bar{U}_2$ subspace non-trivial. The particle-hole symmetric chiral edge modes of a non-trivial model in Hermitian symmetry class C can cross zero energy at an arbitrary edge momentum $k$, which is in general not sampled over in the discrete edge Brillouin zone associated with any finite system size. Hence, there are no exact zero-energy states in the EHH boundary spectrum, and we do not expect to observe an infernal point. The resulting edge states disperse as a function of $k$, forming a NH edge state. Due to the presence of $\bar{U}_{\mathcal{T}}$, which maps between $\bar{U}_1$ subspaces, the entire EHH boundary spectrum hosts two Kramers pairs at symmetric momenta around a TRIM (see Fig.~\ref{fig:sym-class:CII_Spm}a).

For $E_0 \in i\mathbb{R}$, we retain $\bar{U}_{\mathcal{T}}$ [Eq.~\eqref{eq:altTRS}], $\bar{U}_{\mathcal{P}}$ [Eq.~\eqref{eq:altPHS}], $\bar{U}_{\mathcal{C}}, \bar{\Sigma}_{\mathcal{C}}$. This allows to form an additional unitary symmetry $\bar{U}_3 = \bar{U}_{\mathcal{C}} \bar{\Sigma}_{\mathcal{C}}$ with $\bar{U}_3^2 = -1$, which anticommutes with $ \bar{U}_{\mathcal{P}}, \bar{U}_{\mathcal{C}}, \bar{\Sigma}_{\mathcal{C}}$ but commutes with $\bar{U}_{\mathcal{T}}$. Each $\bar{U}_3$ subspace thus possesses the PHS $\bar{U}_{\mathcal{P}}$. The eigensectors of $\bar{U}_3$ are, therefore, not independent, and lie in Hermitian symmetry class C. The pair of chiral modes in one subspace is mapped to a pair of counter-propagating chiral modes through $ \bar{U}_{\mathcal{T}}, \bar{U}_{\mathcal{C}}, \bar{\Sigma}_{\mathcal{C}}$. The two Kramers pairs at symmetric momenta around a TRIM therefore persist, as shown in Fig.~\ref{fig:sym-class:CII_Spm}b.

For $E_0 \in \mathbb{R}$, we are only left with $\bar{U}_1$, whose sectors are mapped onto each other by $\bar{\Sigma}_{\mathcal{C}}, \bar{U}_{\mathcal{T}}$ [Eq.~\eqref{eq:altTRS}] but retain $\bar{U}_{\mathcal{P}}$ [Eq.~\eqref{eq:altPHS}]. This corresponds to Hermitian symmetry class C, with a pair of chiral modes per unitary subspace (see Fig.~\ref{fig:sym-class:CII_Spm}c). 

For arbitrary $E_0 \in \mathbb{C}$, we have $\bar{U}_{\mathcal{T}}$ [Eq.~\eqref{eq:altTRS}], $\bar{U}_{\mathcal{P}}$ [Eq.~\eqref{eq:altPHS}], $\bar{\Sigma}_{\mathcal{C}}$ left, and hence Hermitian symmetry class CI. The two Kramers pairs are no longer protected, removing in-gap states in the EHH boundary spectrum. As a result, the NH edge state disperses as a function of $k$, but only along the real and imaginary axis. This dispersion forms a pair of exceptional points on the edge, represented in the EHH by two Kramers pairs.\newline

\tocless\subsection{NH symmetry class CI$^{S_{-+}}$}{sec:sym-class:CI_Smp}

NH symmetry class CI$^{S_{-+}}$ possesses TRS, PHS and CS with $(U_{\mathcal{T}}U_{\mathcal{T}}^*,U_{\mathcal{P}}U_{\mathcal{P}}^*) = (1,-1)$, where $U_{\mathcal{C}} = U_{\mathcal{T}}U_{\mathcal{P}}$, as well as SLS $\mathcal{S}$. The corresponding EHH at $E_0 =0$ obtains TRS $\bar{U}_{\mathcal{T}}$, PHS $\bar{U}_{\mathcal{P}}$ and three CSs $\bar{U}_{\mathcal{C}}, \bar{\Sigma}_{\mathcal{C}}, \bar{\mathcal{S}}$. The CSs can be combined to two commuting unitary symmetries $\bar{U}_1 = \bar{U}_{\mathcal{C}}\bar{\mathcal{S}}$ with $\bar{U}_1^2 = -1$ and $\bar{U}_2 = \bar{\mathcal{S}} \bar{\Sigma}_{\mathcal{C}}$ with $\bar{U}_2^2 = +1$.

Since $\bar{U}_1$ has an imaginary spectrum and $[\bar{U}_{\mathcal{T}},\bar{U}_1] = \{\bar{U}_{\mathcal{P}},\bar{U}_1\} = \{\bar{U}_{\mathcal{C}},\bar{U}_1\} = \{\bar{\Sigma}_{\mathcal{C}},\bar{U}_1\} = \{\bar{\mathcal{S}},\bar{U}_1\}$, the eigenspaces of $\bar{U}_1$ are not independent and individually enjoy $\bar{U}_{\mathcal{P}}$ and $\bar{U}_2$ symmetry. Moreover, we have $\{\bar{U}_{\mathcal{T}},\bar{U}_2\} = \{\bar{U}_{\mathcal{P}},\bar{U}_2\} = [\bar{\mathcal{S}},\bar{U}_2] = [\bar{U}_{\mathcal{C}},\bar{U}_2] = [\bar{\Sigma}_{\mathcal{C}},\bar{U}_2] =0$, so that the $\bar{U}_2$ eigenspaces are exchanged by $\bar{U}_{\mathcal{P}}$ symmetry. This leaves us with Hermitian symmetry class A, which has a $\z$ classification in 2D~\cite{Ryu2010} that is reduced to $\ztwo$ by line-gap phases~\cite{Sato-PRL:2020}. The non-trivial element corresponds to having a single chiral mode in one $\bar{U}_2$ subspace. The chiral edge mode of a non-trivial model in Hermitian symmetry class A can cross zero energy at an arbitrary edge momentum $k$, which is in general not sampled over in the discrete edge Brillouin zone associated with any finite system size. Hence, there are no exact zero-energy states in the EHH boundary spectrum, and we do not expect to observe an infernal point. The resulting edge state disperses as a function of $k$, forming a NH edge state. Due to the presence of $\bar{U}_{\mathcal{P}}$, which maps between $\bar{U}_2$ subspaces, and $\bar{U}_{\mathcal{T}}$, which maps between $\bar{U}_1$ subspaces, the entire EHH boundary spectrum hosts two Kramers pairs at symmetric momenta around a TRIM (see Fig.~\ref{fig:sym-class:CI_Smp}a).

\begin{figure}
\centering
\includegraphics[width=0.45\textwidth]{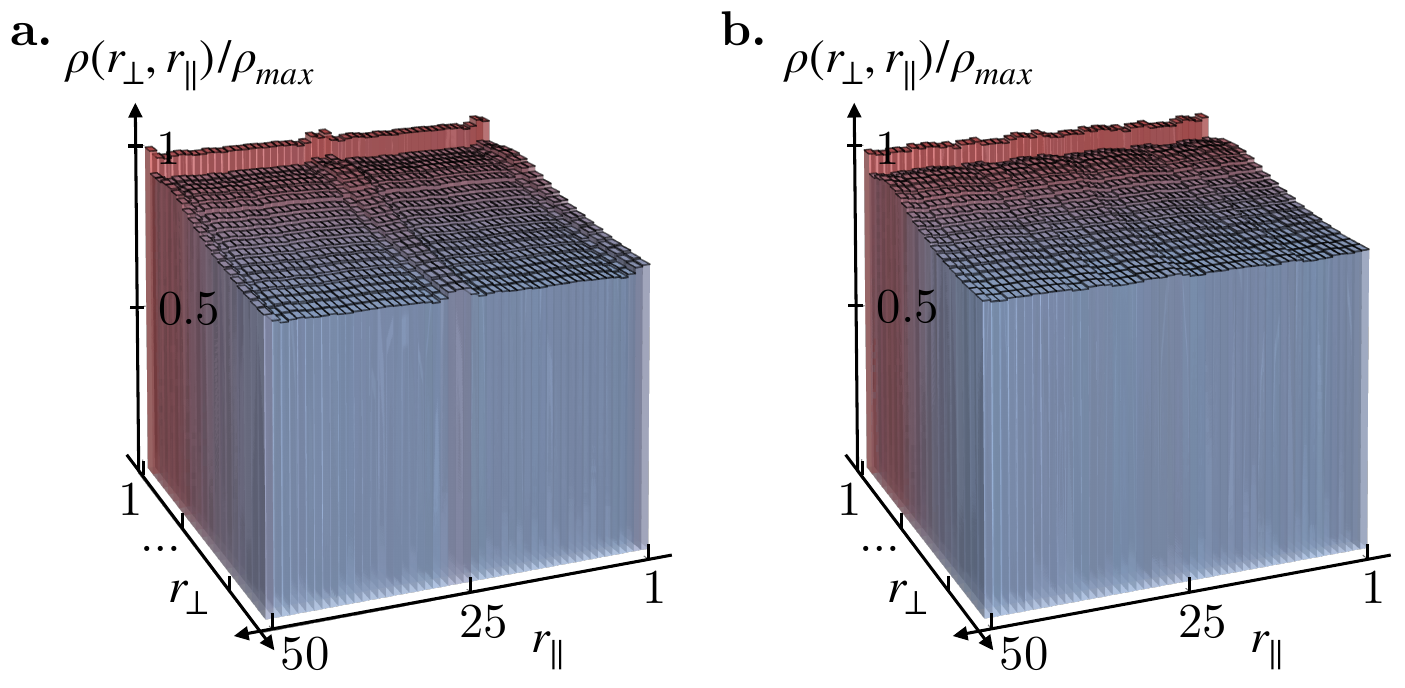}
\caption{\textbf{Exceptional point localization in real space.} \textbf{a.}~Spectrum under OBC in $r_\perp$-direction ($L_\perp = 40$) and PBC in $r_\parallel$-direction ($L_\parallel = 50$). Edge states accumulate at the boundary of the system in $r_\perp$-direction, indicated by a peak of the summed density $\rho(r_\perp, r_\parallel) = \sum_{\alpha, i} |\langle r_\perp, r_\parallel, i | \psi_\alpha\rangle |^2$, where $\alpha$ ranges over \emph{all} eigenstates $\ket{\psi_\alpha}$ of the real-space Hamiltonian, and $i$ runs over all degrees of freedom within the unit cell. \textbf{b.}~Same setting as in panel \textbf{a.}, but in the presence of disorder with $\epsilon = 10^{-6} t$. The exceptional point edge state still prevails, similar to the IP case of Fig.~3c in the main text. 
\label{fig:EPloc}}
\end{figure}
\begin{figure*}
\centering
\includegraphics[width=1\textwidth,page=1]{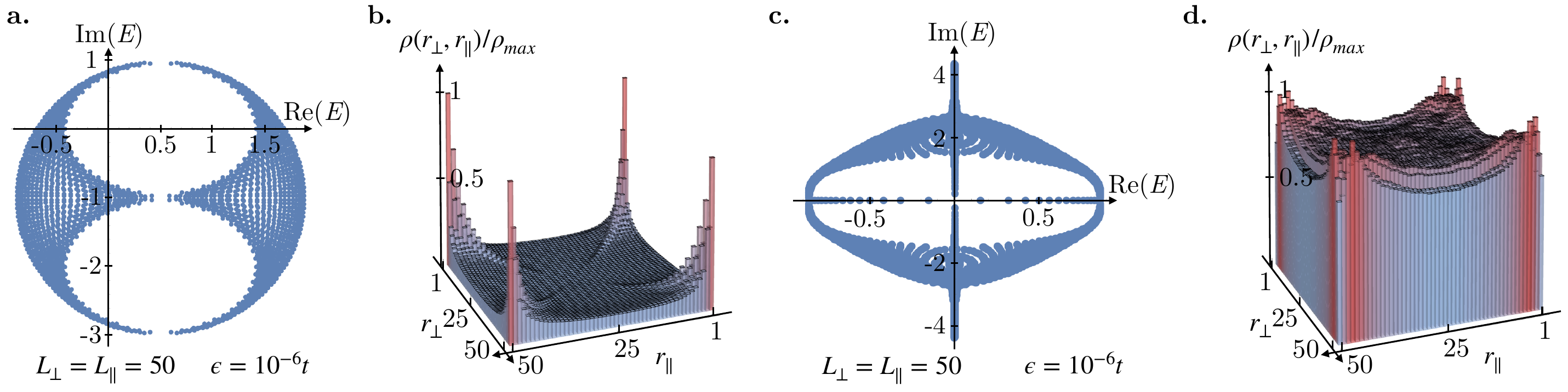}
\caption{\textbf{Full OBC spectrum for infernal and exceptional points.} \textbf{a.}~Spectrum under OBC in both $r_\perp$- and $r_\parallel$-direction ($L_\perp = L_\parallel = 50$) for the infernal point model of Fig.~1 in the main text [Eq.~\eqref{eq:AIIdag}] in presence of disorder with $\epsilon = 10^{-6} t$, where $t$ is the mean hopping strength. In contrast to the SIBC spectrum where the point gap is fully covered by edge-localized states, in-gap states disappear under full OBC. \textbf{b.}~Infernal point localization in real space. In contrary to Fig.~3 of the main text, the boundary does not exhibit an extensive accumulation of eigenmodes along the edges, indicated by the summed density $\rho(r_\perp, r_\parallel) = \sum_{\alpha, i} |\langle r_\perp, r_\parallel, i | \psi_\alpha\rangle |^2$, where $\alpha$ ranges over \emph{all} eigenstates $\ket{\psi_\alpha}$ of the real-space Hamiltonian, and $i$ runs over all degrees of freedom within the unit cell. \textbf{c.}~Spectrum under OBC in both $r_\perp$- and $r_\parallel$-direction ($L_\perp = L_\parallel = 50$) for the exceptional point model of Fig.~2 of the main text [Eq.~\eqref{eq:AIIIS-}] in presence of disorder with $\epsilon = 10^{-6} t$, where $t$ is the mean hopping strength. Unlike for the infernal point, the spectrum retains in-gap states along the real and imaginary axes. \textbf{d.}~Exceptional point localization in real space. %In addition to the spectral signature, surface states remain in the presence of disorder and full OBC.
\label{fig:fullOBC}}
\end{figure*}

For $E_0 \in i\mathbb{R}$, we retain $\bar{U}_{\mathcal{T}}$ [Eq.~\eqref{eq:altTRS}], $\bar{U}_{\mathcal{P}}$ [Eq.~\eqref{eq:altPHS}], $\bar{U}_{\mathcal{C}}, \bar{\Sigma}_{\mathcal{C}}$. This allows to form an additional unitary symmetry $\bar{U}_3 = \bar{U}_{\mathcal{C}} \bar{\Sigma}_{\mathcal{C}}$ with $\bar{U}_3^2 = -1$, which anticommutes with $ \bar{U}_{\mathcal{T}}, \bar{U}_{\mathcal{C}}, \bar{\Sigma}_{\mathcal{C}}$ but commutes with $\bar{U}_{\mathcal{P}}$. Each $\bar{U}_3$ subspace thus possesses the TRS $\bar{U}_{\mathcal{T}}$, resulting in Hermitian symmetry class AII with one Kramers pair. The eigensectors of $\bar{U}_3$ are exchanged by $ \bar{U}_{\mathcal{P}}, \bar{U}_{\mathcal{C}}, \bar{\Sigma}_{\mathcal{C}}$, forming two Kramers pairs at symmetric momenta around a TRIM, as shown in Fig.~\ref{fig:sym-class:CI_Smp}b.

For $E_0 \in \mathbb{R}$, we are only left with $\bar{U}_1$, whose sectors are mapped onto each other by $\bar{\Sigma}_{\mathcal{C}}, \bar{U}_{\mathcal{T}}$ [Eq.~\eqref{eq:altTRS}] but retain $\bar{U}_{\mathcal{P}}$ [Eq.~\eqref{eq:altPHS}]. This corresponds to Hermitian symmetry class D, with two chiral modes in one subspace, mapped to their counter-propagating partners in the other $\bar{U}_1$ subspace. The entire EHH boundary spectrum thus still contains two Kramers pairs at symmetric momenta around a TRIM (see Fig.~\ref{fig:sym-class:CI_Smp}c).

For arbitrary $E_0 \in \mathbb{C}$, we have $\bar{U}_{\mathcal{T}}$ [Eq.~\eqref{eq:altTRS}], $\bar{U}_{\mathcal{P}}$ [Eq.~\eqref{eq:altPHS}], $\bar{\Sigma}_{\mathcal{C}}$ left, and hence Hermitian symmetry class DIII. The two Kramers pairs are no longer protected, removing in-gap states in the EHH boundary spectrum. As a result, the NH edge state disperses as a function of $k$, but only along the real and imaginary axis. This dispersion forms a pair of exceptional points on the edge, represented in the EHH by two Kramers pairs.\newline

\section{Slab solution for infernal points and connection to the $\epsilon$-pseudospectrum}

In order to gain a better understanding of the physical nature of the infernal point, we attempt to solve for the open-boundary conditions (OBC) spectrum of a model in symmetry class AII$^\dagger$. Note that the previous arguments relied on the presence of just a single boundary, i.e. SIBC. Using the analytical approach outlined in Refs.~\onlinecite{Kunst:2020,eti}, we derive an exact slab solution for the specific Hamiltonian introduced in Eq.~\eqref{eq:AIIdag}. Considering OBC in $x-$direction for $L$ unit cells allows to write the Hamiltonian in a block form,
\begin{equation}
    \mathcal{H}_{L,x}(k_y) = \begin{pmatrix}
    \mathcal{H}_{0}(k_y) & \mathcal{H}_{+}  &\cdots & 0 & 0 \\
    \mathcal{H}_{-} & \mathcal{H}_{0}(k_y) & \cdots& 0 & 0 \\
    \vdots & \vdots & \ddots & \vdots & \vdots \\
    0 & 0 & \cdots &\mathcal{H}_{0}(k_y) & \mathcal{H}_{+} \\
    0 & 0 & \cdots &\mathcal{H}_{-} & \mathcal{H}_{0}(k_y)  
    \end{pmatrix}
\end{equation}
with 
\begin{equation}
\mathcal{H}_{0}(k_y) = 
    \begin{pmatrix}
\Delta-i+i\cos k_y & -i\sin k_y\\
i\sin k_y & \Delta -i + i\cos k_y
\end{pmatrix},
\end{equation}
\begin{equation}
\mathcal{H}_+ = 
    \begin{pmatrix}
\frac{i}{2} & -\frac{i}{2}\\
-\frac{i}{2} & \frac{i}{2}
\end{pmatrix},
\end{equation}
\begin{equation}
\mathcal{H}_- = 
    \begin{pmatrix}
\frac{i}{2} & \frac{i}{2}\\
\frac{i}{2} & \frac{i}{2}
\end{pmatrix}.
\end{equation}
To derive the spectrum at $k_y = 0$, we consider the characteristic polynomial $\det[\mathcal{H}_{L,x}(k_y=0)-E \mathbb{1}]$ and solve for its roots. Following the rearrangement of $\mathcal{H}_{L,x}(k_y)$ as a $2\times2$ block matrix as introduced by Refs.~\cite{Kunst:2020,eti}, we can apply Schur's determinant identity and note that 
\begin{equation}
    \mathcal{H}_+ \left[\mathcal{H}_0(k_y=0) -E\mathbb{1}\right]^{-1}\mathcal{H}_- = 0.
\end{equation}
We obtain for $k_y = 0$
\begin{equation}
\begin{aligned}
    \det\left[\mathcal{H}_{L,x}(k_y = 0)-E\mathbb{1}\right] &= \det\left[\mathcal{H}_0(k_y=0)-E\mathbb{1}\right]^L \\&= \left(\Delta -E\right)^{2L}.
\end{aligned}
\end{equation}
This means $E = \Delta$ is a $2L-$fold degenerate solution, which corresponds to the center of the upper point gap in Fig.~1b of the main text. Choosing the lower point gap formed at $k_y = \pi$ yields
\begin{equation}
\begin{aligned}
    \det\left[\mathcal{H}_{L,x}(k_y = \pi)-E\mathbb{1}\right] &= \det\left[\mathcal{H}_0(k_y=\pi)-E\mathbb{1}\right]^L \\&= \left(\Delta -2i-E\right)^{2L}.
\end{aligned}
\end{equation}
corresponding to states in the middle of the lower point gap (see Fig.~1b of the main text), at $E = \Delta-2i$, which are $2L-$fold degenerate. Clearly, this seems to contradict the derivation of the entire point gap being filled by states as outlined in the main text. The reason for this discrepancy lies in the choice of boundary conditions: under OBC, we will always obtain an exponential splitting (in the system size $L$) of the EHH modes on the two edges, thereby not being \emph{exact} zero energy modes. The correspondence with the NH point gap spectrum therefore breaks down, and one does not obtain an edge state at every eigenvalue inside the point gap. To reestablish the NH bulk-boundary correspondence, one should therefore focus on the $\epsilon$-pseudospectrum. It describes the change in the spectrum under a small perturbation $\epsilon$~\cite{pseudospectra,Sato-PRL:2020,Okuma2020},
\begin{equation}
\begin{aligned}
\sigma_\epsilon(H) = \{&E \in \mathbb{C} |~||(H-E) \ket{v}|| < \epsilon\\
&\mathrm{for~at~least~one~}\ket{v}\mathrm{~with~} \braket{v|v}=1\},
\end{aligned}
\end{equation}
which means that in realistic measurements, which always include a small error, a state behaves as an eigenstate if it is just $\mathcal{O}(\epsilon)$ away from being one. The $\epsilon$-pseudospectrum can also be related to the spectrum in SIBC $\sigma_{\mathrm{SIBC}}(H)$, 
\begin{equation}
\lim_{\epsilon \rightarrow 0} \lim_{L \rightarrow \infty} \sigma_\epsilon(H) = \sigma_{\mathrm{SIBC}}(H).
\end{equation}
This correspondence between $\epsilon$-pseudospectrum and SIBC spectrum allows for a precise definition of a NH bulk-boundary correspondence. We investigate this correspondence in Sec.~IV of the main text by numerically introducing on-site disorder terms $V_\epsilon$, which, for fixed system size $L$ and disorder strength $\epsilon$, induce a subset of the full $\epsilon$-pseudospectrum. We find that while spectral signatures differ drastically for infernal and exceptional points, both signatures survive in their respective real space localization (see Fig.~\ref{fig:EPloc}). Figure~\ref{fig:fullOBC} highlights this scenario for full OBC (OBC along two directions), showing the disappearance of infernal point signatures (Fig.~\ref{fig:fullOBC}a,b). Conversely, the exceptional point prevails in the complex spectrum and real space localization (Fig.~\ref{fig:fullOBC}c,d).

\section{Perturbations to the AII$^\dagger$ model}

\begin{figure}
\centering
\includegraphics[width=0.5\textwidth]{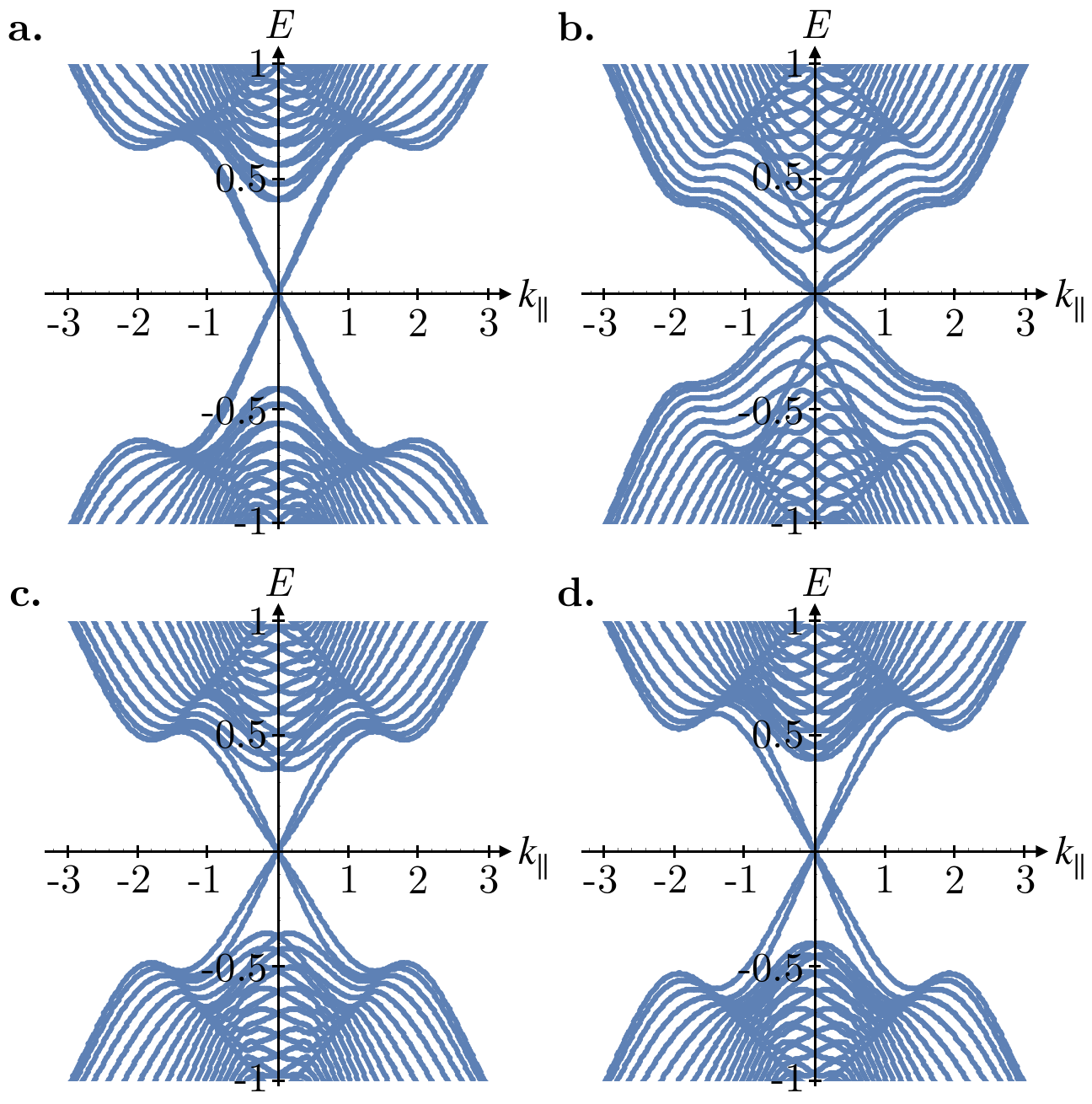}
\caption{\textbf{Stability of the infernal point in class AII$^\dagger$.} In the presence of the perturbations introduced in Eq.~\eqref{eq:AIIdagpert}, the EHH OBC spectrum still exhibits one helical zeromode per edge at $k_\parallel = 0$, for any eigenvalue $E_0$ inside the point gap around $E_0=0$. \textbf{a.}~$(\delta_1,\delta_2,\delta_3) = (0.279734 + 0.0675524 i, 0.110194 + 0.0715397 i, 0.113014 + 0.233133 i)$. \textbf{b.}~$(\delta_1,\delta_2,\delta_3) = (0.211743 + 0.258027 i, 0.265041 + 0.152709 i, 0.141551 + 0.029774 i)$. \textbf{c.}~$(\delta_1,\delta_2,\delta_3) = (0.104291 + 0.0993727 i, 0.19519 + 0.042652 i, 0.125384 + 0.0444993 i)$. \textbf{d.}~$(\delta_1,\delta_2,\delta_3) = (0.121771 + 0.0543309 i, 0.211792 + 0.141878 i, 0.149621 + 0.0950572 i)$.
\label{fig:AIIdagpert}}
\end{figure}

This section investigates the stability of the infernal point of class AII$^\dagger$ under symmetry-allowed perturbations that do not close the point gap and therefore do not change the topological classification of the Hamiltonian of Eq.~\eqref{eq:AIIdag}. Specifically, we consider the most general nearest-neighbor hopping perturbation to our original Hamiltonian that preserves symmetry class AII$^\dagger$:
\begin{equation}
    \begin{aligned}
	\mathcal{H}_{\mathrm{AII}^\dagger}(\bs{k}) + \sum_i \delta_i \sin(k_x) \sigma_i + \sum_i \delta_i \sin(k_y) \sigma_i,
 \label{eq:AIIdagpert}
\end{aligned}
\end{equation}
with the Pauli matrices $\sigma_{i}$ ($i = x, y, z$), possessing TRS$^\dagger$ symmetry in the form of $U_{\mathcal{T}} = i\sigma_y$, and $\delta_i \in \mathbb{C}$ is a complex random prefactor in the range $[0,1]$. The corresponding EHH still shows protected zero-energy modes, pinned to a TRIM (see Fig.~\ref{fig:AIIdagpert}) as a consequence of its symmetry in Altland-Zirnbauer class DIII. In fact, due to time-reversal symmetry of the EHH, symmetry-allowed perturbations that do not close the point gap cannot move zero-energy Kramers pairs away from a TRIM point. This confirms the stability of the infernal point of the corresponding non-Hermitian Hamiltonian and the validity of this result beyond specific model Hamiltonians. For simplicity, we neglect these perturbation terms in the Dirac approximation considered in the main text: Due to the equivalence between the EHH and the corresponding non-Hermitian problem, an infernal point must similarly occur in the Dirac equation obtained by Taylor-expanding Eq.~\eqref{eq:AIIdagpert} to first order in $k$. However, solving this perturbed Dirac equation analytically is much more tedious, and so we have opted to solve the unperturbed equation in the main text for clarity.

\section{Exceptional points in 1D}

This section investigates the stability of exceptional points on the edge of nontrivial 2D NH point-gapped systems.

\tocless\subsection{NH symmetry class AIII$^{S_{-}}$}{sec:EPs-in-AIIISm}

NH symmetry class AIII$^{S_{-}}$ possesses CS $U_{\mathcal{C}}$ as well as SLS $\mathcal{S}$ with $\{U_{\mathcal{C}},\mathcal{S}\} = 0$. The effective Hamiltonain of exceptional points can be written as
\begin{equation}
    \mathcal{H}_{\mathrm{EP}}(k) = \left[\bs{a}(k) + i \bs{b}(k)\right] \cdot \bs{\sigma},
    \label{eq:Hamiltonian-EP}
\end{equation}
with $\bs{a}, \bs{b} \in \mathbb{R}^3$ and $\bs{\sigma} = (\sigma_x, \sigma_y, \sigma_z)^{\mathrm{T}}$ the vector of Pauli matrices. The corresponding spectrum is given by
\begin{equation}
    E_{\mathrm{EP}}(k) = \pm \sqrt{\bs{a}^2(k)-\bs{b}^2(k) + 2i \bs{a}(k) \cdot \bs{b}(k)},
\end{equation}
which allows for a degeneracy if the two conditions
\begin{equation}
    \bs{a}^2(k) = \bs{b}^2(k), \quad \quad \bs{a}(k) \cdot \bs{b}(k)= 0,
    \label{eq:EP-conditions}
\end{equation}
are fulfilled. Choosing $U_{\mathcal{C}} = \sigma_z$ and $\mathcal{S} = \sigma_x$ as a possible representation for NH symmetry class AIII$^{S_{-}}$ restricts $\bs{a}(k) = (0, a_y(k), 0)^{\mathrm{T}}$ and $\bs{b}(k) = (0, 0, b_z(k))^{\mathrm{T}}$ in Eq.~\eqref{eq:Hamiltonian-EP}. Correspondingly, the second condition in Eq.~\eqref{eq:EP-conditions} is automatically fulfilled, leaving us with a single constraint~\cite{Kawabata:2019:3,Sayyad:2022,Sayyad:2022:2}. This can be realized with a single momentum, allowing for exceptional points in the edge Brillouin zone of NH symmetry class AIII$^{S_{-}}$.

\tocless\subsection{NH symmetry class DIII$^{S_{+-}}$}{sec:EPs-in-DIIISpm}

NH symmetry class DIII$^{S_{+-}}$ possesses TRS, PHS and CS with $(U_{\mathcal{T}}U_{\mathcal{T}}^*,U_{\mathcal{P}}U_{\mathcal{P}}^*) = (-1,1)$, where $U_{\mathcal{C}} = U_{\mathcal{T}}U_{\mathcal{P}}$, as well as SLS $\mathcal{S}$. One specific choice fulfilling this constraint is given by $U_{\mathcal{T}} = \tau_0 \sigma_y$, $U_{\mathcal{P}} = \tau_x \sigma_x$, $U_{\mathcal{C}} = \tau_x \sigma_z$, $\mathcal{S} = \tau_z \sigma_0$ with the Pauli matrices $\tau_\mu, \sigma_\mu$, $\mu = 0,x,y,z$. To satisfy the constraints of CS and SLS, the effective Hamiltonain of exceptional points in NH symmetry class DIII$^{S_{+-}}$ can be written as
\begin{equation}
\begin{aligned}
    \mathcal{H}_{\mathrm{EP}}(k) =&\, a_1(k) \tau_x \sigma_x +a_2(k) \tau_x \sigma_y\\
    &+ a_3(k) \tau_y \sigma_0+ a_4(k) \tau_y \sigma_z\\
    &+i b_1(k) \tau_x \sigma_0 +i b_2(k) \tau_x \sigma_z \\
    &+ i b_3(k) \tau_y \sigma_x+ i b_4(k) \tau_y \sigma_z,
\end{aligned}
\end{equation}
with $a_i, b_i \in \mathbb{R}$, $i = 1,\ldots, 4$. The resulting complex spectrum follows as 
\begin{equation}
    E_{\mathrm{EP}}(k) = \pm \sqrt{f_1(a_i,b_i;k) \pm \sqrt{f_2(a_i,b_i;k)}},
\end{equation}
where $f_{1,2}$ are functions of $a_i, b_i$, $i = 1,\ldots, 4$. For an exceptional point with a two-fold degeneracy, only a single condition $|f_1(a_i,b_i;k)| = |\sqrt{f_2(a_i,b_i;k)}|$ needs to be fulfilled~\cite{Kawabata:2019:3,Sayyad:2022,Sayyad:2022:2}. This is possible with a single momentum, allowing for exceptional points in the edge Brillouin zone of NH symmetry class DIII$^{S_{+-}}$. At a TRIM, the presence of TRS and PHS imposes further constraints on the Hamiltonian,
\begin{equation}
    \mathcal{H}_{\mathrm{EP}}(k \in \mathrm{TRIM}) =  a_4 \tau_y \sigma_z +i b_2 \tau_x \sigma_z,
\end{equation}
where we omitted the momentum dependence in $a_4, b_2$ for simplicity. The corresponding spectrum results in a four-fold degenerate exceptional point for the fine-tuned case of $a_4(k \in \mathrm{TRIM}) = b_2(k \in \mathrm{TRIM})$. As outlined in Sec.~\ref{sec:sym-class:DIIISpm}, this degeneracy is stable, since each exceptional point corresponds to a zero-energy Kramers pair in distinct unitary subspaces of the EHH. 

\tocless\subsection{NH symmetry class CI$^{S_{-+}}$}{sec:EPs-in-CISmp}

NH symmetry class CI$^{S_{-+}}$ possesses TRS, PHS and CS with $(U_{\mathcal{T}}U_{\mathcal{T}}^*,U_{\mathcal{P}}U_{\mathcal{P}}^*) = (1,-1)$, where $U_{\mathcal{C}} = U_{\mathcal{T}}U_{\mathcal{P}}$, as well as SLS $\mathcal{S}$. One specific choice fulfilling this constraint is given by $U_{\mathcal{T}} = \tau_y \sigma_y$, $U_{\mathcal{P}} = \tau_z \sigma_y$, $U_{\mathcal{C}} = \tau_x \sigma_0$, $\mathcal{S} = \tau_z \sigma_0$ with the Pauli matrices $\tau_\mu, \sigma_\mu$, $\mu = 0,x,y,z$. To satisfy the constraints of CS and SLS, the effective Hamiltonain of exceptional points in NH symmetry class CI$^{S_{-+}}$ can be written as
\begin{equation}
\begin{aligned}
    \mathcal{H}_{\mathrm{EP}}(k) =&\, a_1(k) \tau_y \sigma_0 +a_2(k) \tau_y \sigma_x\\
    &+ a_3(k) \tau_y \sigma_y+ a_4(k) \tau_y \sigma_z\\
    &+i b_1(k) \tau_x \sigma_0 +i b_2(k) \tau_x \sigma_x \\
    &+ i b_3(k) \tau_x \sigma_y+ i b_4(k) \tau_x \sigma_z,
\end{aligned}
\end{equation}
with $a_i, b_i \in \mathbb{R}$, $i = 1,\ldots, 4$. The resulting complex spectrum follows as 
\begin{equation}
    E_{\mathrm{EP}}(k) = \pm \sqrt{f_1(a_i,b_i;k) \pm \sqrt{f_2(a_i,b_i;k)}},
\end{equation}
where $f_{1,2}$ are functions of $a_i, b_i$, $i = 1,\ldots, 4$. For an exceptional point with a two-fold degeneracy, only a single condition $|f_1(a_i,b_i;k)| = |\sqrt{f_2(a_i,b_i;k)}|$ needs to be fulfilled~\cite{Kawabata:2019:3,Sayyad:2022,Sayyad:2022:2}. This is possible with a single momentum, allowing for exceptional points in the edge Brillouin zone of NH symmetry class CI$^{S_{-+}}$. At a TRIM, the presence of TRS and PHS imposes further constraints on the Hamiltonian,
\begin{equation}
\begin{aligned}
    \mathcal{H}_{\mathrm{EP}}(k \in \mathrm{TRIM}) =&\,  a_2 \tau_y \sigma_x+ a_3 \tau_y \sigma_y\\
    &+ a_4 \tau_y \sigma_z +i b_1 \tau_x \sigma_0,
    \end{aligned}
\end{equation}
where we omitted the momentum dependence in $a_{2,3,4}, b_1$ for simplicity. The corresponding spectrum results in a four-fold degenerate exceptional point for the fine-tuned case of 
\begin{equation}
\begin{aligned}
    b_1^2 (k \in \mathrm{TRIM})= & \, a_2^2 (k \in \mathrm{TRIM})\\
    & + a_3^2 (k \in \mathrm{TRIM})\\
    &+ a_4^2 (k \in \mathrm{TRIM}).
    \end{aligned}
\end{equation}
As outlined in Sec.~\ref{sec:sym-class:CI_Smp}, this degeneracy is stable, since each exceptional point corresponds to a zero-energy Kramers pair in distinct unitary subspaces of the EHH. 

\tocless\subsection{NH symmetry class CII$^{S_{-+}}$}{sec:EPs-in-CIISpm}

NH symmetry class CII$^{S_{-+}}$ possesses TRS, PHS and CS with $(U_{\mathcal{T}}U_{\mathcal{T}}^*,U_{\mathcal{P}}U_{\mathcal{P}}^*) = (-1,-1)$, where $U_{\mathcal{C}} = U_{\mathcal{T}}U_{\mathcal{P}}$, as well as SLS $\mathcal{S}$. One specific choice fulfilling this constraint is given by $U_{\mathcal{T}} = \tau_x \sigma_y$, $U_{\mathcal{P}} = \tau_0 \sigma_y$, $U_{\mathcal{C}} = \tau_x \sigma_0$, $\mathcal{S} = \tau_z \sigma_0$ with the Pauli matrices $\tau_\mu, \sigma_\mu$, $\mu = 0,x,y,z$. To satisfy the constraints of CS and SLS, the effective Hamiltonain of exceptional points in NH symmetry class CII$^{S_{-+}}$ can be written as
\begin{equation}
\begin{aligned}
    \mathcal{H}_{\mathrm{EP}}(k) =&\, a_1(k) \tau_y \sigma_0 +a_2(k) \tau_y \sigma_x\\
    &+ a_3(k) \tau_y \sigma_y+ a_4(k) \tau_y \sigma_z\\
    &+i b_1(k) \tau_x \sigma_0 +i b_2(k) \tau_x \sigma_x \\
    &+ i b_3(k) \tau_x \sigma_y+ i b_4(k) \tau_x \sigma_z,
\end{aligned}
\end{equation}
with $a_i, b_i \in \mathbb{R}$, $i = 1,\ldots, 4$. The resulting complex spectrum follows as 
\begin{equation}
    E_{\mathrm{EP}}(k) = \pm \sqrt{f_1(a_i,b_i;k) \pm \sqrt{f_2(a_i,b_i;k)}},
\end{equation}
where $f_{1,2}$ are functions of $a_i, b_i$, $i = 1,\ldots, 4$. For an exceptional point with a two-fold degeneracy, only a single condition $|f_1(a_i,b_i;k)| = |\sqrt{f_2(a_i,b_i;k)}|$ needs to be fulfilled~\cite{Kawabata:2019:3,Sayyad:2022,Sayyad:2022:2}. This is possible with a single momentum, allowing for exceptional points in the edge Brillouin zone of NH symmetry class CI$^{S_{-+}}$. At a TRIM, the presence of TRS and PHS imposes further constraints on the Hamiltonian,
\begin{equation}
\begin{aligned}
    \mathcal{H}_{\mathrm{EP}}(k \in \mathrm{TRIM}) =&\,  a_1 \tau_y \sigma_0+ i b_2 \tau_x \sigma_x\\
    &+ i b_3 \tau_x \sigma_y +i b_4 \tau_x \sigma_z,
    \end{aligned}
\end{equation}
where we omitted the momentum dependence in $a_1, b_{2,3,4}$ for simplicity. The corresponding spectrum results in a four-fold degenerate exceptional point for the fine-tuned case of 
\begin{equation}
\begin{aligned}
    a_1^2 (k \in \mathrm{TRIM})= & \, b_2^2 (k \in \mathrm{TRIM})\\
    & + b_3^2 (k \in \mathrm{TRIM})\\
    &+ b_4^2 (k \in \mathrm{TRIM}).
    \end{aligned}
\end{equation}
As outlined in Sec.~\ref{sec:sym-class:CII_Smp}, this degeneracy is stable, since each exceptional point corresponds to a pair of zero-energy states in distinct unitary subspaces of the EHH. 

\tocless\subsection{NH symmetry class CII$^{S_{+-}}$}{sec:EPs-in-CIISmp}

NH symmetry class CII$^{S_{+-}}$ possesses TRS, PHS and CS with $(U_{\mathcal{T}}U_{\mathcal{T}}^*,U_{\mathcal{P}}U_{\mathcal{P}}^*) = (-1,-1)$, where $U_{\mathcal{C}} = U_{\mathcal{T}}U_{\mathcal{P}}$, as well as SLS $\mathcal{S}$. One specific choice fulfilling this constraint is given by $U_{\mathcal{T}} = \tau_0 \sigma_y$, $U_{\mathcal{P}} = \tau_y \sigma_0$, $U_{\mathcal{C}} = \tau_y \sigma_y$, $\mathcal{S} = \tau_x \sigma_0$ with the Pauli matrices $\tau_\mu, \sigma_\mu$, $\mu = 0,x,y,z$. To satisfy the constraints of CS and SLS, the effective Hamiltonain of exceptional points in NH symmetry class CII$^{S_{+-}}$ can be written as
\begin{equation}
\begin{aligned}
    \mathcal{H}_{\mathrm{EP}}(k) =&\, a_1(k) \tau_y \sigma_x +a_2(k) \tau_y \sigma_z\\
    &+ a_3(k) \tau_z \sigma_0+ a_4(k) \tau_z \sigma_y\\
    &+i b_1(k) \tau_y \sigma_0 +i b_2(k) \tau_y \sigma_y \\
    &+ i b_3(k) \tau_z \sigma_x+ i b_4(k) \tau_z \sigma_z,
\end{aligned}
\end{equation}
with $a_i, b_i \in \mathbb{R}$, $i = 1,\ldots, 4$. The resulting complex spectrum follows as 
\begin{equation}
    E_{\mathrm{EP}}(k) = \pm \sqrt{f_1(a_i,b_i;k) \pm \sqrt{f_2(a_i,b_i;k)}},
\end{equation}
where $f_{1,2}$ are functions of $a_i, b_i$, $i = 1,\ldots, 4$. For an exceptional point with a two-fold degeneracy, only a single condition $|f_1(a_i,b_i;k)| = |\sqrt{f_2(a_i,b_i;k)}|$ needs to be fulfilled~\cite{Kawabata:2019:3,Sayyad:2022,Sayyad:2022:2}. This is possible with a single momentum, allowing for exceptional points in the edge Brillouin zone of NH symmetry class CII$^{S_{+-}}$. At a TRIM, the presence of TRS and PHS imposes further constraints on the Hamiltonian,
\begin{equation}
\begin{aligned}
    \mathcal{H}_{\mathrm{EP}}(k\in \mathrm{TRIM}) =&\, a_1 \tau_y \sigma_x +a_2 \tau_y \sigma_z\\
    &+ a_3 \tau_z \sigma_0+i b_1 \tau_y \sigma_0\\
    &+ i b_3 \tau_z \sigma_x+ i b_4 \tau_z \sigma_z,
\end{aligned}
\end{equation}
where we omitted the momentum dependence in $a_{1,2,3}, b_{1,3,4}$ for simplicity. The corresponding spectrum results in a four-fold degenerate exceptional point for the fine-tuned case of 
\begin{equation}
    a_1^2 + a_2^2 + a_3^2 = b_1^2 + b_3^2 + b_4^2
\end{equation}
and
\begin{equation}
\begin{aligned}
    0 =&-a_2^2 (b_1^2 + b_3^2) - 2 a_2  a_3  b_1  b_4 + 2 a_1  b_3 ( a_2  b_4-a_3  b_1)\\
    &- a_1^2 (b_1^2 + b_4^2) - a_3^2 (b_3^2 + b_4^2)
\end{aligned}
\end{equation}
at a TRIM. As outlined in Sec.~\ref{sec:sym-class:CII_Spm}, this degeneracy is stable, since each exceptional point corresponds to a pair of zero-energy states in distinct unitary subspaces of the EHH. 

\section{Toy model Hamiltonians}

In this section, we present models for all intrinsically point-gapped nontrivial 2D NH symmetry classes.

\tocless\subsection{NH symmetry class AII$^\dagger$}{sec:AIIdag-model}

The model in NH symmetry class AII$^\dagger$ is given by the Hamiltonian~\cite{fluxes}
\begin{equation}
	\label{eq:AIIdag}
	\begin{aligned}
	\mathcal{H}_{\mathrm{AII}^\dagger}(\bs{k}) =& \sin(k_x) \sigma_x - \sin(k_y) \sigma_z\\ &+ i  \left(\sum_{i = x,y} \cos(k_i)-1 \right) \sigma_0 + \Delta \sigma_0,
\end{aligned}
\end{equation}
with the Pauli matrices $\sigma_{\mu}$ ($\mu = 0, x, y, z$), possessing TRS$^\dagger$ symmetry in the form of $U_{\mathcal{T}} = i\sigma_y$ and $\Delta \neq 0$ is a real parameter breaking residual symmetries. For Fig.~1 in the main text, we use $\Delta = 0.5$.

\tocless\subsection{NH symmetry class DIII$^\dagger$}{sec:DIIIdag-model}

The model in NH symmetry class DIII$^\dagger$ is given by the Hamiltonian~\cite{fluxes}
\begin{equation}
	\label{eq:DIIIdag}
	\begin{aligned}
	\mathcal{H}_{\mathrm{DIII}^\dagger}(\bs{k}) =& \sin(k_x) \sigma_y + \sin(k_y) \sigma_x\\ &+ i  \left(\sum_{i = x,y} \cos(k_i)-1 \right) \sigma_0,
\end{aligned}
\end{equation}
with the Pauli matrices $\sigma_{\mu}$ ($\mu = 0, x, y, z$), possessing pseudo TRS as $U_{\mathcal{T}} = i\sigma_y$, pseudo PHS as $U_{\mathcal{P}} = \sigma_x$, and CS as $U_{\mathcal{C}} = \sigma_z$.

\tocless\subsection{NH symmetry class BDI$^{S_{+-}}$}{sec:BDISpm-model}

The nontrivial point-gapped model in NH symmetry class BDI$^{S_{+-}}$ is realized by the Hamiltonian 
\begin{equation}
	\label{eq:BDIS+-}
	\begin{aligned}
	\mathcal{H}_{\mathrm{BDI}^{S_{+-}}}(\bs{k}) =\begin{pmatrix}
		0 & Q(\bs{k}; 1) \\
		Q(\bs{k}; -1) & 0\\
	\end{pmatrix},
	\end{aligned}
\end{equation}
with 
\begin{equation}
\begin{aligned}
	Q(\bs{k};\mu)=& i\sin(k_x) \sigma_x +  i\sin(k_y) \sigma_y\\ &+ i\left(\sum_{i = x,y} \cos(k_i)-2+\mu \right) \sigma_z,
	\end{aligned}
\end{equation}
with the Pauli matrices $\sigma_{\mu}$ and $\tau_{\mu}$ ($\mu = 0, x, y, z$).
The Hamiltonian possesses PHS $U_{\mathcal{P}} = \tau_x \sigma_x$, TRS $U_{\mathcal{T}} = \tau_0 \sigma_x$,  CS $U_{\mathcal{C}} = \tau_x \sigma_0$ and sublattice symmetry $\mathcal{S} = \tau_z \sigma_0$, fulfilling the required commutation relations.

\tocless\subsection{NH symmetry class D$^{S_{-}}$}{sec:DSm-model}

A nontrivial point-gapped model in NH symmetry class D$^{S_{-}}$ is realized by the Hamiltonian 
\begin{equation}
	\label{eq:DS-}
	\begin{aligned}
	\mathcal{H}_{\mathrm{D}^{S_-}}(\bs{k}) = \begin{pmatrix}
		0 & Q(\bs{k}; 1) \\
		Q(\bs{k}; 3) & 0\\
	\end{pmatrix} + \Delta \tau_x \sigma_y,
	\end{aligned}
\end{equation}
with 
\begin{equation}
\begin{aligned}
	Q(\bs{k};\mu)=& i \sin(k_x) \sigma_z +  \sin(k_y) \sigma_0\\ &+ 2i\left( \sum_{i = x,y} \cos(k_i)-\mu \right) \sigma_y,
	\end{aligned}
\end{equation}
with the Pauli matrices $\sigma_{\mu}$ and $\tau_{\mu}$ ($\mu = 0, x, y, z$).
The Hamiltonian possesses PHS $U_{\mathcal{P}} = \tau_x \sigma_0$ and sublattice symmetry $\mathcal{S} = \tau_z \sigma_0$, fulfilling the required commutation relations. Note that $\Delta$ multiplies a term to remove unwanted residual symmetries.

\tocless\subsection{NH symmetry class AIII$^{S_{-}}$}{sec:AIIISm-model}

The nontrivial point-gapped model in NH symmetry class AIII$^{S_{-}}$ is realized by the Hamiltonian 
\begin{equation}
	\label{eq:AIIIS-}
	\begin{aligned}
	\mathcal{H}_{\mathrm{AIII}^{S_-}}(\bs{k}) = \begin{pmatrix}
		0 & Q(\bs{k}; 1) \\
		Q(\bs{k}; -1) & 0\\
	\end{pmatrix},
	\end{aligned}
\end{equation}
with 
\begin{equation}
\begin{aligned}
	Q(\bs{k};\mu)=& i\sin(k_x) \sigma_x +  i\sin(k_y) \sigma_y\\ &+ i\left(\sum_{i = x,y} \cos(k_i)-2+\mu \right) \sigma_z + i \Delta \sigma_0,
	\end{aligned}
\end{equation}
with the Pauli matrices $\sigma_{\mu}$ and $\tau_{\mu}$ ($\mu = 0, x, y, z$).
The Hamiltonian possesses CS $U_{\mathcal{C}} = \tau_x \sigma_0$ and sublattice symmetry $\mathcal{S} = \tau_z \sigma_0$, fulfilling the required commutation relations. Note that $\Delta$ multiplies a term to remove unwanted residual symmetries, which we select as $\Delta = 0.3$ in Fig.~2 and $\Delta = 0.5$ in Fig.~3 of the main text.

\tocless\subsection{NH symmetry class DIII$^{S_{+-}}$}{sec:DIIISpm-model}

The nontrivial point-gapped model in NH symmetry class DIII$^{S_{+-}}$ is realized by the Hamiltonian 
\begin{equation}
	\label{eq:DIIIS+-}
	\begin{aligned}
	\mathcal{H}_{\mathrm{DIII}^{S_{+-}}}(\bs{k}) = \begin{pmatrix}
		0 & Q(\bs{k}; -3) \\
		Q(\bs{k}; -1) & 0\\
	\end{pmatrix},
	\end{aligned}
\end{equation}
with 
\begin{equation}
\begin{aligned}
	Q(\bs{k};\mu)=& i \left(\sum_{i = x,y} \cos(k_i) + \mu \right) \tau_x \sigma_z\\
 &+ i \sin(k_x) \tau_0\sigma_x +  \sin(k_y) \tau_z\sigma_x\\ 
 &+ i \Delta \left( \tau_0 \sigma_y -i  \tau_z \sigma_y +i  \tau_y \sigma_y\right),
	\end{aligned}
\end{equation}
with the Pauli matrices $\sigma_{\mu}$, $\tau_{\mu}$, and  $\rho_{\mu}$ ($\mu = 0, x, y, z$). The Hamiltonian possesses PHS $U_{\mathcal{P}} = \rho_x \tau_z \sigma_0$, TRS $U_{\mathcal{T}} = \rho_0 \tau_y \sigma_0$,  CS $U_{\mathcal{C}} =  \rho_x \tau_x \sigma_0$ and sublattice symmetry $\mathcal{S} = \rho_z \tau_0 \sigma_0$, fulfilling the required commutation relations.

\tocless\subsection{NH symmetry class CII$^{S_{-+}}$}{sec:CIISmp-model}

The nontrivial point-gapped model in NH symmetry class CII$^{S_{-+}}$ is realized by the Hamiltonian 
\begin{equation}
	\label{eq:CIIS-+}
	\begin{aligned}
	\mathcal{H}_{\mathrm{CII}^{S_{-+}}}(\bs{k}) = \begin{pmatrix}
		0 & Q(\bs{k}) \\
		-\sigma_x Q(\bs{k})^* \sigma_x & 0\\
	\end{pmatrix},
	\end{aligned}
\end{equation}
with 
\begin{equation}
\begin{aligned}
	Q(\bs{k})=& i \left(\sum_{i = x,y} \cos(k_i) -1 \right) \sigma_z\\
 &+ i \sin(k_x) \sigma_x +  i\sin(k_y) \sigma_y+ i\Delta \sigma_0,
	\end{aligned}
\end{equation}
with the Pauli matrices $\sigma_{\mu}$ and $\tau_{\mu}$ ($\mu = 0, x, y, z$). The Hamiltonian possesses PHS $U_{\mathcal{P}} = \tau_0 \sigma_y$, TRS $U_{\mathcal{T}} = \tau_x \sigma_y$,  CS $U_{\mathcal{C}} = \tau_x \sigma_0$ and sublattice symmetry $\mathcal{S} = \tau_z \sigma_0$, fulfilling the required commutation relations.

\tocless\subsection{NH symmetry class CII$^{S_{+-}}$}{sec:CIISpm-model}

The nontrivial point-gapped model in NH symmetry class CII$^{S_{+-}}$ is realized by the Hamiltonian 
\begin{equation}
	\label{eq:CIIS+-}
	\begin{aligned}
	\mathcal{H}_{\mathrm{CII}^{S_{+-}}}(\bs{k}) = \begin{pmatrix}
		0 & Q(\bs{k}; -3) \\
		Q(\bs{k}; -1) & 0\\
	\end{pmatrix},
	\end{aligned}
\end{equation}
with 
\begin{equation}
\begin{aligned}
	Q(\bs{k};\mu)=& i \left(\sum_{i = x,y} \cos(k_i) + \mu \right) \tau_z \sigma_x\\
 &+ i \sin(k_x) \tau_x\sigma_x +  i\sin(k_y) \tau_y\sigma_0\\ 
 &+ \Delta \sum_{i = x,y,z} \tau_0 \sigma_i,
	\end{aligned}
\end{equation}
with the Pauli matrices $\sigma_{\mu}$, $\tau_{\mu}$, and  $\rho_{\mu}$ ($\mu = 0, x, y, z$). The Hamiltonian possesses PHS $U_{\mathcal{P}} = \rho_x \tau_z \sigma_y$, TRS $U_{\mathcal{T}} = \rho_0 \tau_z \sigma_y$,  CS $U_{\mathcal{C}} =  \rho_x \tau_0 \sigma_0$ and sublattice symmetry $\mathcal{S} = \rho_z \tau_0 \sigma_0$, fulfilling the required commutation relations. 

\tocless\subsection{NH symmetry class CI$^{S_{-+}}$}{sec:CISmp-model}

The nontrivial point-gapped model in NH symmetry class CI$^{S_{-+}}$ is realized by the Hamiltonian 
\begin{equation}
	\label{eq:CIS-+}
	\begin{aligned}
	\mathcal{H}_{\mathrm{CI}^{S_{-+}}}(\bs{k}) = \begin{pmatrix}
		0 & Q(\bs{k}) \\
		\sigma_x Q(\bs{k})^* \sigma_x & 0\\
	\end{pmatrix},
	\end{aligned}
\end{equation}
with 
\begin{equation}
\begin{aligned}
	Q(\bs{k})=& i \left(\sum_{i = x,y} \cos(k_i) -1 \right) \sigma_z\\
 &+ i \sin(k_x) \sigma_x +  i\sin(k_y) \sigma_y+ i\Delta \sigma_0,
	\end{aligned}
\end{equation}
with the Pauli matrices $\sigma_{\mu}$ and $\tau_{\mu}$ ($\mu = 0, x, y, z$). The Hamiltonian possesses PHS $U_{\mathcal{P}} = \tau_z \sigma_y$, TRS $U_{\mathcal{T}} = \tau_y \sigma_y$,  CS $U_{\mathcal{C}} = \tau_x \sigma_0$ and sublattice symmetry $\mathcal{S} = \tau_z \sigma_0$, fulfilling the required commutation relations. 

\let\oldaddcontentsline\addcontentsline% Store \addcontentsline
\renewcommand{\addcontentsline}[3]{}% Make \addcontentsline a no-op
\bibliography{bibliography}
\let\addcontentsline\oldaddcontentsline% Restore \addcontentsline